\documentclass[12pt]{article}
\pdfoutput=1
\usepackage{amsmath,amssymb,graphicx,epsfig,color,float,cancel,cite}
\usepackage[titletoc,title]{appendix} 
\usepackage[export]{adjustbox}
\usepackage{multicol}
\usepackage{hyperref}                                                                                                                                                                                                                                                                                                                                                                                                                                                                                                                                                                                                                                                                                                                                                                                                                                                                                                                                                                                                                                                                                                        
\numberwithin{equation}{section}

\def\nn{\nonumber}
\def\GeV{{\, \rm GeV}}
\def\TeV{{\, \rm TeV}}
\def\fb{{\, \rm fb}}
\def\pb{{\, \rm pb}}
\def\nn{\nonumber}
\def\bea{\begin{eqnarray}}
\def\eea{\end{eqnarray}}

\def\fvev{\langle F_x \rangle}
\def\svev{\langle {\cal S} \rangle}
\def\SU#1{ {\cal SU}(#1)}

\newcommand{\beq}{\begin{equation}}
\newcommand{\eeq}{\end{equation}}
\newcommand{\bseq}{\begin{subequations}}
\newcommand{\eseq}{\end{subequations}}
\newcommand{\beqa}{\begin{eqnarray}}
\newcommand{\eeqa}{\end{eqnarray}}
\begin{document}
\thispagestyle{empty}
\begin{center}
\vspace*{1cm}
{\LARGE\bf The 750 GeV diphoton resonance as an sgoldstino: a reappraisal\\}
\bigskip
{\large Debjyoti Bardhan}\,$^{a,1}$, \, \, 
{\large Pritibhajan Byakti}\,$^{b,2}$,\\
{\large Diptimoy Ghosh}\,$^{c,3}$,\, \, 
{\large Tarun Sharma}\,$^{c,4}$
\\
\bigskip 
{\small
$^a$ Department of Theoretical Physics, Tata Institute of Fundamental 
Research, \\ 1 Homi Bhabha Road, Mumbai 400005, India. 

$^b$ Center for High Energy Physics, Indian Institute of Science, 
\\ Bangalore 560012, India.

$^c$ Department of Particle Physics and Astrophysics, Weizmann Institute of Science, 
\\ Rehovot 76100, Israel. 
}
\end{center}
\vspace*{1cm}
\begin{center} 
{\Large\bf Abstract} 
\end{center}
\vspace*{-0.35in}
\begin{quotation}
\noindent  Among the various explanations of the possible 750 GeV diphoton resonance,  the possibility of it being 
an sgoldstino is an attractive one, as it is related to the spontaneous breaking of global supersymmetry. We 
discuss this possibility in this paper and point out the various theoretical issues associated with it. In particular, 
we indicate the difficulties of this explanation in realistic models of gauge mediated supersymmetry breaking. 
\end{quotation}
\bigskip
%
%
\vfill
%
%
\bigskip
\hrule
\vspace*{-0.1in}
$^1$ debjyoti@theory.tifr.res.in \hfill
$^2$ pritibhajan@cts.iisc.ernet.in \\
$^3$ diptimoy.ghosh@weizmann.ac.il\hfill
$^4$ tarun.sharma@weizmann.ac.il

\newpage 
\tableofcontents
\section{Introduction}
The ATLAS and the CMS collaborations have recently reported some excess of events in the diphoton invariant mass 
($m_{\gamma\gamma}$) distribution based on 3.2 and 2.6 fb$^{-1}$ of proton-proton collision data respectively collected at a 
center-of-mass energy of 13 TeV. ATLAS observed the most significant deviation from the background hypothesis 
at $m_{\gamma\gamma} \approx 750$ GeV, corresponding to a local (global) significance of 3.6 (2.0)
\footnote{This was obtained using a narrow width of the signal component. The statistical significance increases slightly once the possibility of 
larger width is taken into account. See \cite{ATLAS-CONF-2015-081} for more details.} \cite{ATLAS-CONF-2015-081}.  
The largest excess in the CMS data was seen around the 760 GeV mass bin with a local (global) significance of \mbox{2.6 ($\lesssim 1.2$)} 
standard deviations \cite{CMS-PAS-EXO-15-004}. 
This excess is also found consistent with the constraints from the run 1 data \cite{Falkowski:2015swt}. 
It was also reported by ATLAS that the  properties of the events in the signal region  were found to be compatible with those in the invariant 
mass regions above and below the excess.  
As suggested by many authors, the most simple-minded explanation of this excess is to propose the existence of a resonance ($\cal S$)
of mass $\sim 750 \GeV$. In order to generate the correct amount of signal, the resonance must have couplings that produce 
$\sigma^{\rm signal} \equiv \sigma(p p \to {\cal S}) {\rm Br}({\cal S} \to \gamma \, \gamma)$ about 5\fb \cite{Gupta:2015zzs,Aloni:2015mxa,Falkowski:2015swt}.

In this article, we consider the possibility of this resonance being an sgoldstino\footnote{To our knowledge, the name ``sgoldstino" was 
first used in \cite{Brignole:1998uu}.}, the ``superpartner" 
of the goldstino, the goldstone fermion of spontaneous global supersymmetry (SUSY) breaking. 
This possibility has been discussed by \cite{Bellazzini:2015nxw,Petersson:2015mkr,Demidov:2015zqn,Casas:2015blx,Ding:2016udc} 
using an effective description of how the SUSY breaking is mediated to the MSSM sector. 
In this article, we scrutinise the viability of this proposal when realistic models for the mediation of 
SUSY breaking are considered.  But before we start discussing that, we would like to make a few general comments about SUSY 
breaking in order to put things in perspective.

Unlike other symmetries, there are some interesting limitations on the possibility of spontaneous global SUSY breaking. 
For example, neither a pure super Yang-Mills (SYM) nor a SYM theory with massive matter in real representations of the 
gauge group breaks SUSY spontaneously\footnote{This follows from the fact that Witten Index of these theories is 
non-zero \cite{Witten:1982df}. See also \cite{Izawa:1996pk,Intriligator:1996pu}.}. 
In particular, global  N = 2 SYM theories (that have matter in real representations) cannot have SUSY spontaneously broken. 
This is one of the reasons why one needs global N = 1 SUSY with complex representation for phenomenology (i.e., MSSM) as there 
is a hope that SUSY can be spontaneously broken as required by experiments.

However, even in MSSM, it turns out to be impossible to break SUSY spontaneously. In fact, with the minimal field content of 
MSSM both the SUSY and the EW symmetry remain unbroken\footnote{A Fayet-Iliopoulos $D$-term breaking also turns out to be 
phenomenologically unacceptable \cite{Martin:1997ns}.}. Hence, adding more fields to the MSSM is unavoidable. 
However, even after adding many heavy fields, the gaugino masses cannot arise in a renormalisable SUSY theory at tree-level. 
This is because SUSY does not contain any (gaugino)-(gaugino)-(scalar) coupling that could give rise to a gaugino mass term when 
the scalar gets a vacuum expectation value (VEV). 
Moreover, the tree level supertrace rules do not allow a phenomenologically acceptable spectrum. 

Hence, one possibility for breaking SUSY spontaneously in the MSSM is to have tree level SUSY breaking in a so-called ``hidden sector" 
and radiatively mediate the information of SUSY breaking to the MSSM sector\footnote{Note that, in four space-time dimensions, 
if supersymmetry is not  broken spontaneously at the tree level, then it can not be  broken by radiative Coleman-Weinberg 
mechanism \cite{Gates:1983nr}}. 
This also helps in finding a solution of the SUSY flavour problem. As the pattern of SUSY breaking interactions 
in the visible MSSM sector is determined by the interactions of the messenger particles with the MSSM, a natural way to avoid 
additional flavour violation in the MSSM is to have flavour symmetries in the messenger interactions. 
The models of gauge mediation, where the information of SUSY breaking is communicated to the MSSM sector by gauge interactions, 
achieve this goal in a natural way\footnote{This is however not true in general, as the messenger fields can have renormalisable 
superpotential couplings to the MSSM \cite{Han:1998xy,Shadmi:2011hs,Kang:2012ra,Craig:2012xp,Byakti:2013ti,Evans:2013kxa}}.  
In the gauge mediation scenarios, one assumes the existence of ``messenger fields" that are charged both under the SM gauge group 
as well as the hidden sector quantum numbers. The mass scale of these messengers is arbitrary and, in principle, can be as low as $\sim 10 \TeV$. 
These models are often called ``low scale SUSY breaking" scenarios and, as we 
will see later, are the  only ones 
(among the different SUSY breaking scenarios) relevant for the diphoton excess.

In the following section, we review the general framework that leads to the sgoldstino explanation of the diphoton excess and present 
the necessary formulae to study the phenomenology. In section~\ref{OGM}, we will discuss the ordinary gauge mediation (OGM) scenario 
and point out the various theoretical issues it confronts in connection to the diphoton excess. The generalisation of the 
OGM framework, called the extraordinary gauge mediation (EOGM), will be discussed in section~\ref{EOGM}. In section~\ref{way-out}, 
we will investigate whether there is some way out of the difficulties raised in the previous sections. We will conclude in section~\ref{conclusion}.

\section{Generalities}
\label{general}

\subsection{Theoretical framework}

In order to parameterise the effect of SUSY breaking in the visible sector, it is usually assumed that SUSY is broken in
the hidden sector by the VEV of the $F$ component of a chiral superfield $X$. In particular, the 
gaugino masses are generated by the following terms, 
\begin{eqnarray}
\label{eff-Lag-1}
\mathcal{L} \subset 
-\frac{1}{2} \frac{c_1}{M_1} \int d^2\theta X W^{1 \, \alpha} W_\alpha^1 
&-&\frac{1}{2} \frac{c_2}{M_2} \int d^2\theta X W^{2 \, \alpha \, A} W_\alpha^{2 \, A} \nn \\
&-&\frac{1}{2} \frac{c_3}{M_3} \int d^2\theta X W^{3 \, \alpha \, A} W_\alpha^{3 \, A} + {\rm h.c.}
\end{eqnarray}
where the superscripts \{1,2,3\} refer to the $U(1)$, $SU(2)$ and $SU(3)$ gauge groups respectively (the adjoint indices for both the 
gauge groups $SU(2)$ and $SU(3)$ are denoted by $A$), and $\alpha$ is the spinor index.  The scale $M_i$ denotes the mass scale 
of the messeger fields which have been integrated out to get the above Lagrangian terms\footnote{In models of gravity mediation, 
the scale $M_i$ is of the order of the planck scale. It is then clear that gravity mediation models are not relevant for the diphoton excess.}. 
The chiral superfield $X$ and $W_\alpha$ 
have the following expansion in terms of the ordinary fields,
\begin{eqnarray}
X &=& {\cal S}  + \sqrt{2} \theta \xi(y) + \theta \theta F_x(y) \\
&=& \frac{1}{\sqrt{2}} \left(\phi(y) + i a(y) \right) + \sqrt{2} \theta \xi(y) + \theta \theta F_x(y) \\
W_\alpha^{A} &=&  -i\lambda_\alpha^A (y) + D^A(y) \theta_\alpha - (\sigma^{\mu \nu} \theta)_\alpha F_{\mu\nu}^A(y) - \theta\theta\sigma^\mu_{\alpha \dot{\beta}}D^{(y)}_\mu \lambda^{\dagger \, A \, \dot{\beta}}(y) \, ,
\end{eqnarray}
where, $y^\mu = x^\mu - i \theta \sigma^\mu \theta^\dagger$.

Once the $F$ term of $X$ gets a VEV, say $\langle F_x \rangle$, the above Lagrangian terms generate the following Majorana masses for 
the gauginos,
\begin{eqnarray}
\label{c-couplings}
m_i &=& c_i \,  \dfrac{\langle F_x \rangle}{M_i} \, .
\end{eqnarray}

The Lagrangian of Eq.~\eqref{eff-Lag-1} also generates couplings of the scalar components of $X$ to the gauge bosons, 
\begin{eqnarray}
\mathcal{L}_{gg}    &=&  \frac{1}{2 \sqrt{2}}  \, \, \frac{c_3}{M_3}  \, \left(\phi G_{\mu\nu}^a G^{\mu\nu a} - a G_{\mu\nu} \tilde{G}^{\mu\nu}\right) \label{Lgg}\\
\mathcal{L}_{WW} &=&   \frac{1}{2\sqrt{2}} \, \,  \frac{2 c_2}{M_2} \, \left(\phi W^{+}_{\mu \nu}W^{- \mu \nu} - a W^+_{\mu\nu} \tilde{W}^{-\mu\nu}\right)  \label{Lww}\\
\mathcal{L}_{\gamma \gamma} &=& \frac{1}{2 \sqrt{2}} \, \, \left(\frac{c_1}{M_1} c_W^2 + \frac{c_2}{M_2} s_W^2 \right) \, \left(\phi F_{\mu \nu} F^{\mu \nu} - 
a F_{\mu\nu} \tilde{F}^{\mu\nu}\right) \label{Lgaga}\\
\mathcal{L}_{ZZ} &=& \frac{1}{2 \sqrt{2}} \, \, \left(\frac{c_1}{M_1} s_W^2 + \frac{c_2}{M_2} c_W^2 \right)   \, \left(\phi Z_{\mu \nu} Z^{\mu \nu} - 
a Z_{\mu\nu} \tilde{Z}^{\mu\nu}\right) \label{Lzz}\\
\mathcal{L}_{Z\gamma} &=& \frac{1}{2\sqrt{2}} \, \,  2 s_W c_W \, \left(\frac{c_2}{M_2} - \frac{c_1}{M_1} \right)  \, \left(\phi Z_{\mu \nu} F^{\mu \nu} - 
a Z_{\mu\nu} \tilde{F}^{\mu\nu}\right) \label{Lzga} \, .
\end{eqnarray}

The scalars $\phi$ and $a$ can decay to the gauge bosons through these couplings. The corresponding partial decay rates are given by
(see appendix~\ref{app-B} for details)\footnote{Signatures of sgoldstino at the $e^+ e^-$ and hadron colliders were first 
studied in \cite{Perazzi:2000id,Perazzi:2000ty} where the formulae for the decay rates can also be found.} 
\begin{eqnarray}
\Gamma_{\gamma\gamma} \equiv \Gamma (\phi \to \gamma \gamma) &=& 
\left[\dfrac{1}{2 m_\phi}\right]  \left[\dfrac{1}{8\pi}\right]  \left[\dfrac{1}{8}\left(\frac{c_1}{M_1} c_W^2 + \frac{c_2}{M_2} s_W^2 \right)^2\right]  
\left[8m_\phi^4\right] \left[\dfrac{1}{2}\right]\\
\Gamma_{gg} \equiv \Gamma (\phi \to gg) &=&\left[\dfrac{1}{2 m_\phi}\right]  \left[\dfrac{1}{8\pi}\right] \left[ \frac{1}{8} \left(\frac{c_3}{M_3}\right)^2\right]  
\left[ 64 m_\phi^4\right] \left[ \frac{1}{2}\right] \\
\Gamma_{z\gamma} \equiv \Gamma (\phi \to Z \gamma) &=& \left[\dfrac{1}{2 m_\phi}\right] \left[\dfrac{1}{8\pi}\left(1 - \frac{m_Z^2}{m_\phi^2}\right)\right] \, 
\left[\dfrac{1}{8}  \left(\frac{c_2}{M_2} - \frac{c_1}{M_1} \right)^2 4 s_W^2 c_W^2\right] \\
&& \times \left[2 m_\phi^4 \left(1 - \frac{m_Z^2}{m_\phi^2}\right)^2\right]  \nn \\
\Gamma_{zz} \equiv \Gamma (\phi \to ZZ) &=& \left[\dfrac{1}{2 m_\phi}\right] \left[\dfrac{1}{8\pi}\left(1 - 4\frac{m_Z^2}{m_\phi^2}\right)^{1/2}\right] \, 
\left[\dfrac{1}{8}  \left(\frac{c_1}{M_1} s_W^2 + \frac{c_2}{M_2} c_W^2 \right)^2\right] \nn \\
&& \times \left[8 m_\phi^4 \left(1 - 4\frac{m_Z^2}{m_\phi^2} + 6\frac{m_Z^4}{m_\phi^4} \right)\right] \left[ \frac{1}{2}\right] 
\eea
\bea
\Gamma_{ww} \equiv \Gamma (\phi \to WW) &=& \left[\dfrac{1}{2 m_\phi}\right] \left[\dfrac{1}{8\pi}\left(1 - 4\frac{m_W^2}{m_\phi^2}\right)^{1/2}\right] \, 
\left[\dfrac{1}{8}  \left(\frac{2 c_2}{M_2} \right)^2\right] \nn \\
&& \times \left[8 m_\phi^4 \left(1 - 4\frac{m_W^2}{m_\phi^2} + 6\frac{m_W^4}{m_\phi^4} \right)\right]
\end{eqnarray}
Here $s_W$ and $c_W$ denote the sine and cosine of the Weinberg angle respectively. The partial decay rates for the scalar $a$ can be obtained 
from the above expressions by replacing $m_\phi$ by $m_a$. There is slight difference between the decay rates of $\phi \to ZZ(W^+W^-)$ and 
$a \to ZZ(W^+W^-)$; however, that is numerically insignificant (see appendix~\ref{app-B}).

\subsection{Explaining the excess}
\label{excess}

The total cross section for the diphoton production via the resonance ${\cal S}$ is given by\footnote{Here we use the approximation 
that $\Gamma_{{\cal S}}/m_{{\cal S}}$ is small. This is a very good approximation even for the case when $\Gamma=45 \GeV$, which gives 
$\Gamma_{{\cal S}}/m_{{\cal S}} = 0.06$.}, 
\begin{eqnarray}
\sigma_{_{\rm LHC\,energy}} &=& \sigma(p \, p \to {\cal S})_{{\rm LHC\,energy}} \, {\rm Br}({\cal S} \to \gamma \, \gamma) \nn \\
&=& \sum_i {\mathcal A}^{ii}_{{\rm LHC\,energy}} \, \Gamma({\cal S} \to p_i \, p_i) \, \dfrac{ \Gamma({\cal S} \to \gamma \, \gamma)}{\Gamma_{{\cal S}}} \, ,
\end{eqnarray}
where $\{p_i \, p_i\}$ refers to the initial state partons i.e., $\{g \,g\}$,  $\{\bar{u}\,u\}$, $\{\bar{d}\,d\}$ and so on. The total width of 
${\cal S}$ is denoted by $\Gamma_{{\cal S}}$.
The numerical values of the quantities ${\mathcal A}^{ii}_{{\rm LHC\,energy}}$ are calculated in appendix~\ref{app-A} and are given 
by,

\begin{multicols}{2}
\noindent
\begin{align}
{\mathcal A}_{13}^{gg} \equiv {\mathcal A}|_{{\rm 13\,TeV\,LHC}}^{gg} &=&  \dfrac{ 5.44 \pb}{\GeV}      \nn \\
{\mathcal A}_{13}^{\bar{u}u} \equiv {\mathcal A}|_{{\rm 13\,TeV\,LHC}}^{\bar{u}u} &=&  \dfrac{ 2.94 \pb}{\GeV}  \nn \\
{\mathcal A}_{13}^{\bar{d}d} \equiv {\mathcal A}|_{{\rm 13\,TeV\,LHC}}^{\bar{d}d} &=&  \dfrac{ 1.73 \pb}{\GeV} \nn 
\end{align}
\begin{align}
{\mathcal A}_{8}^{gg} \equiv {\mathcal A}|_{{\rm 8\,TeV\,LHC}}^{gg} &=& \dfrac{ 1.15 \pb}{\GeV}    \nn \\
{\mathcal A}_{8}^{\bar{u}u} \equiv {\mathcal A}|_{{\rm 8\,TeV\,LHC}}^{\bar{u}u} &=&  \dfrac{1.2 \pb}{\GeV} \\
{\mathcal A}_{8}^{\bar{d}d} \equiv {\mathcal A}|_{{\rm 8\,TeV\,LHC}}^{\bar{d}d} &=&  \dfrac{ 0.66 \pb}{\GeV} \nn 
\end{align}
\end{multicols}

In order to explain the signal, $\sigma_{13\TeV}$ must be approximately in the range $3 - 8 \fb$, assuming that the resonance 
has a small width $\lesssim \textnormal{few} \GeV$ \cite{Falkowski:2015swt}. A larger width of $\sim~40~\GeV$ requires $\sigma_{13\TeV}$ 
to be slightly higher: $\sigma_{13\TeV} \approx 5 -14 \fb$ \cite{Falkowski:2015swt}. As the sgoldstino typically has a narrow width, in our 
estimates we will use the range $3 - 8 \fb$ for the required cross section.

We will first consider the production by gluon fusion only, as the production by $u\bar{u}$ and $d\bar{d}$ initial states is slightly 
disfavoured \cite{Gupta:2015zzs,Aloni:2015mxa,Falkowski:2015swt}. In section~\ref{qqbar-production}, we will comment on the possibility 
of quark initiated production. 

\section{Ordinary gauge mediation}
\label{OGM}

In the OGM framework, the hidden sector is parameterised by a single chiral superfield $X$.  Both the scalar 
and auxiliary components of $X$ are assumed to get VEVs that are denoted by $\langle {\cal S} \rangle$ and $\langle F_x \rangle$ respectively. 
In addition to this, OGM also includes $N_5$ vector like pairs of messenger fields, ($\Phi_i, \tilde \Phi_i$), transforming under 
${\bf 5} + {\bf \bar{5}}$ of $SU(5)$\footnote{Complete representations of a GUT group are normally used in order to keep the unification 
of the gauge couplings intact. However, in general, complete representations are not necessary.  The use of incomplete representations often  
also have interesting phenomenology, see for example, \cite{Byakti:2012qk} and the references therein.}. 
The corresponding superpotential reads, 
\begin{eqnarray}
\label{OGM-Lag}
W_{{\rm OGM}} = \lambda_{ij} X \tilde \Phi_i \Phi_j \, ,
\end{eqnarray}
where the indices $\{i,j\}$ run from 1 to $N_5$. Note that the matrix $\lambda_{ij}$ can always be brought to a diagonal form with real entries by 
independent unitary rotations on $\Phi$ and  $\tilde \Phi$ (the K{\"a}hler potential remain unchanged). Hence, in the rest of this section, we will 
assume that $\lambda_{ij}$ is diagonal with $\lambda_{ii} \equiv \lambda_{i}$.  

The fermions of each $\{\Phi_i, \tilde \Phi_i\}$ pair has a Dirac mass  $m_F^i = \lambda_{i} \langle {\cal S} \rangle$. The mass eigenstates of the 
complex scalars, on the other hand, have squared masses  $m_\pm^{i2} = \lambda_{i}^2 \langle {\cal S} \rangle^2 \pm \lambda_i 
\langle F_x \rangle$. The gaugino masses are generated at the one loop level and are given by
\cite{Giudice:1998bp},
\begin{eqnarray}
\label{OGM-gaugino-mass}
m_a = \frac{\alpha_a}{4 \pi}  \sum_{i=1}^{N_5}  d_i^a \,  \frac{\lambda_i \fvev}{m_F^i} \, g(x_i)   \qquad  (a = 1,2,3)
\end{eqnarray}
where, 
$x_i = \dfrac{\lambda_i \fvev}{(m_F^i)^2}$ and the function $g(x)$ is given by\cite{Giudice:1998bp}, 
\bea
g(x) = \frac{1}{x^2}[(1+x){\rm Log}(1+x) + (1-x){\rm Log}(1-x)] \, .
\eea
The symbol $d_i$ denotes twice the Dynkin index for a particular representation. For example, 
in the case of ${\bf 5} + {\bf \bar{5}}$ of $SU(5)$, $d = 1$. In Eq.~\ref{OGM-gaugino-mass}, we have used the GUT normalisation of the 
hypercharge gauge coupling.

Note that the SUSY breaking $F$-term VEV $\fvev$ must satisfy $\fvev \leq \lambda_i \svev^2  \, ,\forall i$ in order to avoid the messenger 
scalar masses from becoming tachyonic. For simplicity, we assume all the $\lambda_i$ couplings to be equal and set them to a common value $\lambda$. 
We define the ratio $\lambda \fvev/ m_F^2$ to be $\kappa$. With these definitions, the formula for the gaugino mass takes the form (for messengers 
in ${\bf 5} + {\bf \bar{5}}$ of $SU(5)$),
\begin{eqnarray}
\label{OGM-gaugino-mass-2}
m_a = \frac{\alpha_a}{4 \pi} \, \kappa \, m_F \, N_5 \, g(\kappa)   \qquad  (a = 1,2,3) \, .
\end{eqnarray}
The $c_a$ couplings (see Eq.~\ref{c-couplings}) which control the signal strength are given by,  
\bea
\frac{c_a}{M_a} = \frac{m_a}{\fvev} =  \frac{\alpha_a}{4 \pi} \, \frac{\lambda}{m_F} \, N_5 \, g(\kappa)   \qquad  (a = 1,2,3) \, .
\label{eq:ci}
\eea

Similarly, the scalar masses can be written as \cite{Dimopoulos:1996gy,Martin:1996zb}, 
\bea
{\widetilde m}^2_a &=& 2 N_5 \, \kappa^2 \, m_F^2  
\left[C_3^a \left(\frac{\alpha_3}{4 \pi}\right)^2  + C_2^a \left(\frac{\alpha_2}{4 \pi}\right)^2 + C_1^a \left(\frac{\alpha_1}{4 \pi}\right)^2 \right] f(\kappa)
\eea
where $C^a$ are the quadratic Casimirs and the function $f(x)$ is given by\cite{Giudice:1998bp}, 
\bea
f(x) = \frac{1+x}{x^2}\left[{\rm Log}(1+x) - 2 {\rm Li}_2\left(\frac{x}{1+x}\right) + \frac{1}{2} {\rm Li}_2\left(\frac{2x}{1+x}\right)\right] + (x \to -x) \, .
\eea

In order to calculate the gaugino masses at the $\sim \TeV$ scale, we use the values of $\alpha_a$ at $1 \TeV$, which we compute using 
the one loop SM running equations,
\begin{eqnarray}
\frac{1}{\alpha_a (\mu)} &\simeq& \frac{1}{\alpha_a (m_Z)} + \frac{b_a}{\pi} \, {\rm Ln}\left(\frac{\mu}{m_Z}\right) \nn \\
\{\frac{1}{\alpha_1 (m_Z)}, \frac{1}{\alpha_2 (m_Z)}, \frac{1}{\alpha_3 (m_Z)}\} &=& \{ 59, 30, 8.5\} \\
\{b_1^{\rm SM},b_2^{\rm SM},b_3^{\rm SM}\} &=& \{- \frac{41}{20},\frac{19}{12},\frac{7}{2}\}
\end{eqnarray}

We now examine the requirements on $m_F$, $\fvev$ and $N_5$ in order to generate the correct cross section for the excess. 
In order to have a feeling for the messenger mass scale required for the excess, we first consider a single pair of $SU(5)$ messengers 
$\{{\bf 5} + \overline{{\bf 5}}\}$ i.e., $N_5 =1$ and also set $\lambda =1$.  
Following the discussion of the previous section, the explanation of the diphoton excess requires\footnote{Here we have neglected 
any decay mode other than the gauge boson final states. However, existence of other decay modes will increase the total width of the resonance, 
hence adding an extra contribution to the denominator of Eq.~\ref{feeling}. This means that the required 
signal cross section will be even higher, as pointed out also in the end of section~\ref{excess}. Thus our estimate is on the 
conservative side.}, 
\begin{eqnarray}
\label{feeling}
{\mathcal A}_{13}^{gg} \, \frac{\Gamma_{gg} \Gamma_{\gamma\gamma}}
{\Gamma_{gg} + \Gamma_{\gamma\gamma} + \Gamma_{ww} + \Gamma_{zz}+\Gamma_{z\gamma}} \gtrsim 3 \fb \, .
\end{eqnarray}
This gives,
\bea
m_F \lesssim  175 \GeV \, .
\eea

\begin{figure}[t]
\centering
\begin{tabular}{c c }
\includegraphics[scale=0.8]{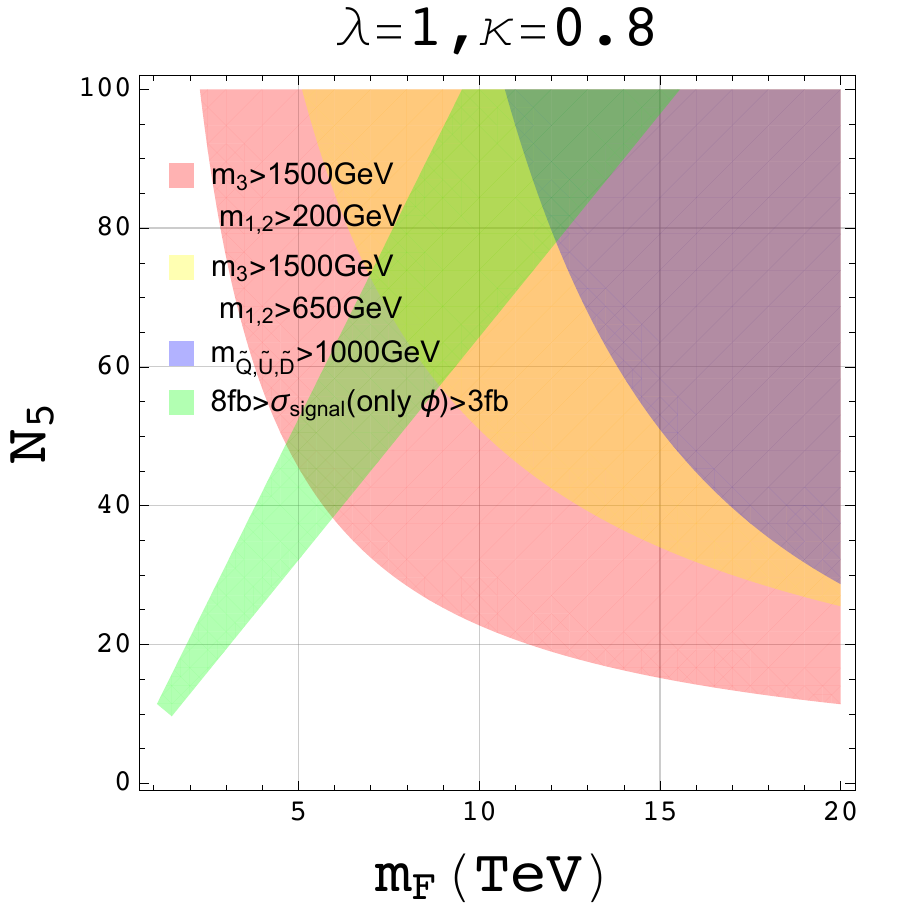} & ~~~~~\includegraphics[scale=0.8]{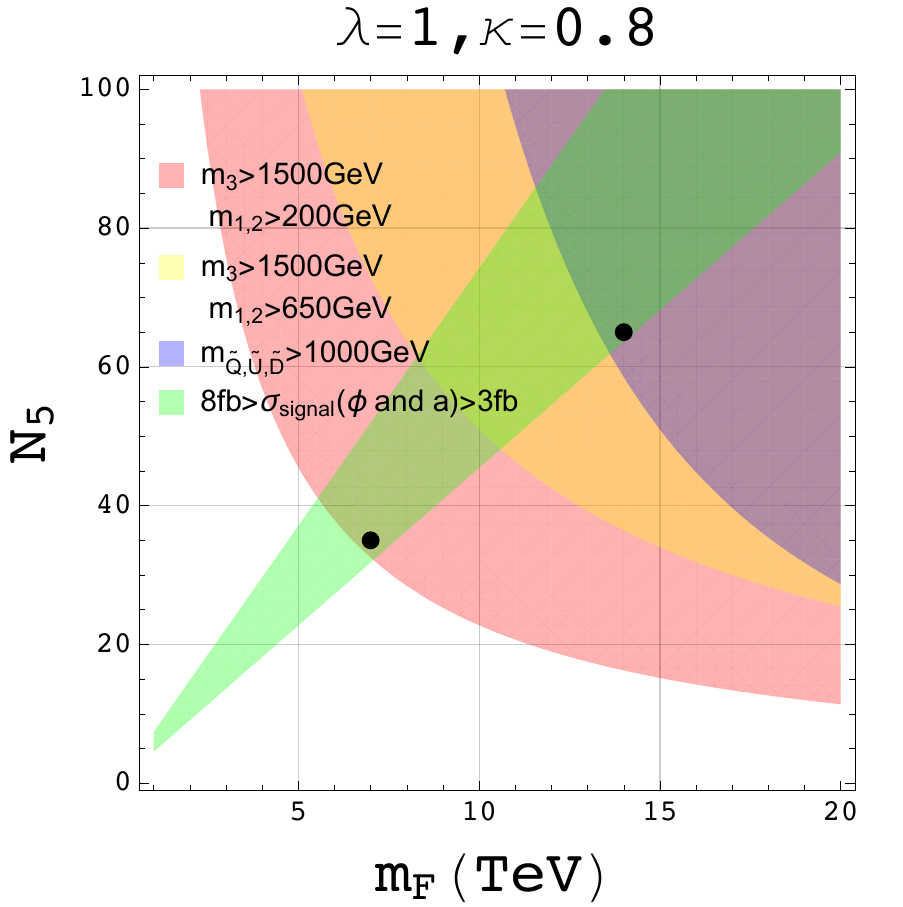}
\end{tabular}
\caption{{\sf Allowed region in the OGM parameter space that successfully explains the signal and satisfy LHC bounds on squark and gaugino masses. While in the left panel the contribution from only $\phi$ is considered, the right panel takes into account both $\phi$ and $a$ contributions.}}
\label{fig:OGM-param}
\end{figure}

The messenger scale can be raised if the number of messenger fields is increased. In Fig.~\ref{fig:OGM-param} 
we show the allowed region in the $m_F$ -- $N_5$ plane for $\lambda =1$ and $\kappa=0.8$. In the left panel, only the contribution 
of $\phi$ to the signal is considered, while in the upper right panel contributions from both $\phi$ and $a$ are taken into account. 
As discussed before, $\kappa$ should satisfy  
$\kappa \leq 1$ to avoid tachyonic states in the messenger sector. For $\kappa$ very close to unity, one of the complex scalars in every pair of 
messenger fields becomes too light (its squared mass is $m_F^2 (1-\kappa)$).  Also, the function $f(\kappa)$ decreases rapidly for 
$\kappa \gtrsim 0.8$ \cite{Giudice:1998bp} reducing the MSSM squark masses. Hence, we have chosen a value $\kappa = 0.8$ in 
Fig.~\ref{fig:OGM-param}.

The light green shaded region reproduces the correct amount of signal to explain the excess. In the light red shaded region, the gaugino 
masses are what is required by the exclusion limits of the LHC. In particular, the gluino mass is set to more than 1.5 TeV and 
a conservative lower bound of 200 GeV is considered for the bino and wino masses 
(we also show the region satisfying a stricter lower bound of 650 GeV on the bino and wino masses \cite{Aad:2015hea}). Similarly, in the 
light blue region the squarks are heavier than a $\TeV$. It can be seen that a very large number of messengers $\gtrsim 60$ is required 
in order to both successfully explain the signal as well as produce sufficiently large gaugino and squark masses.

\begin{figure}[t]
\centering
\begin{tabular}{c c }
\includegraphics[scale=0.8]{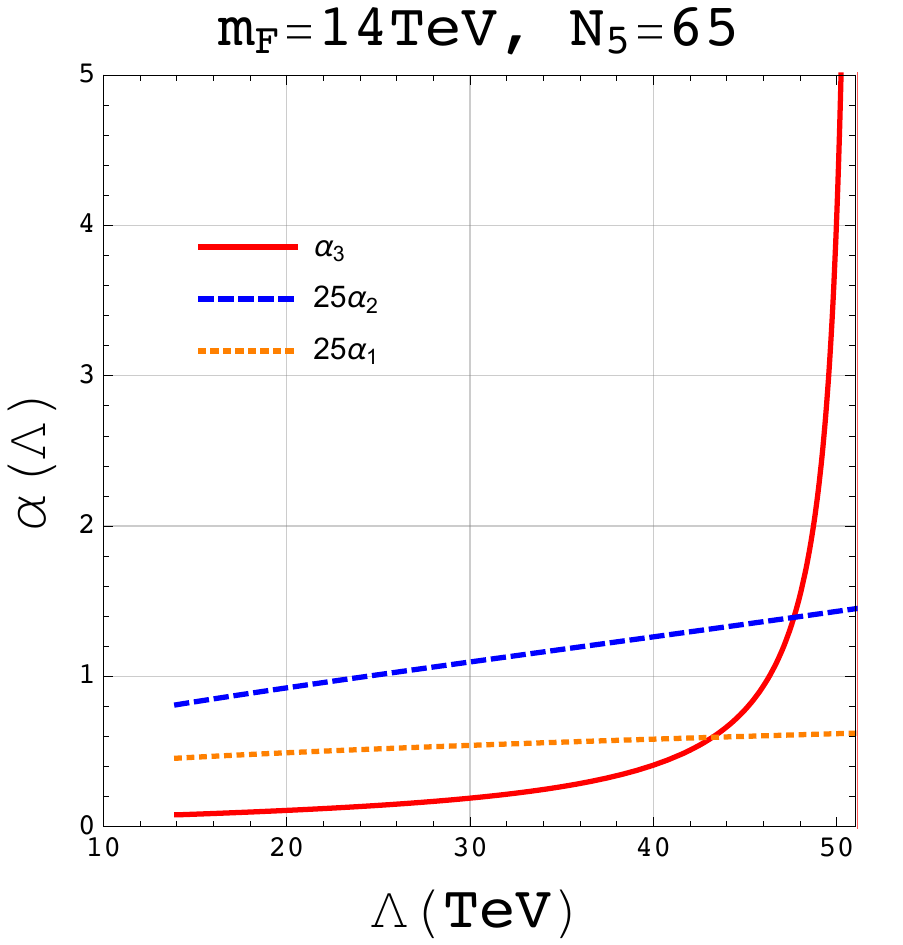} & ~~~~ \includegraphics[scale=0.8]{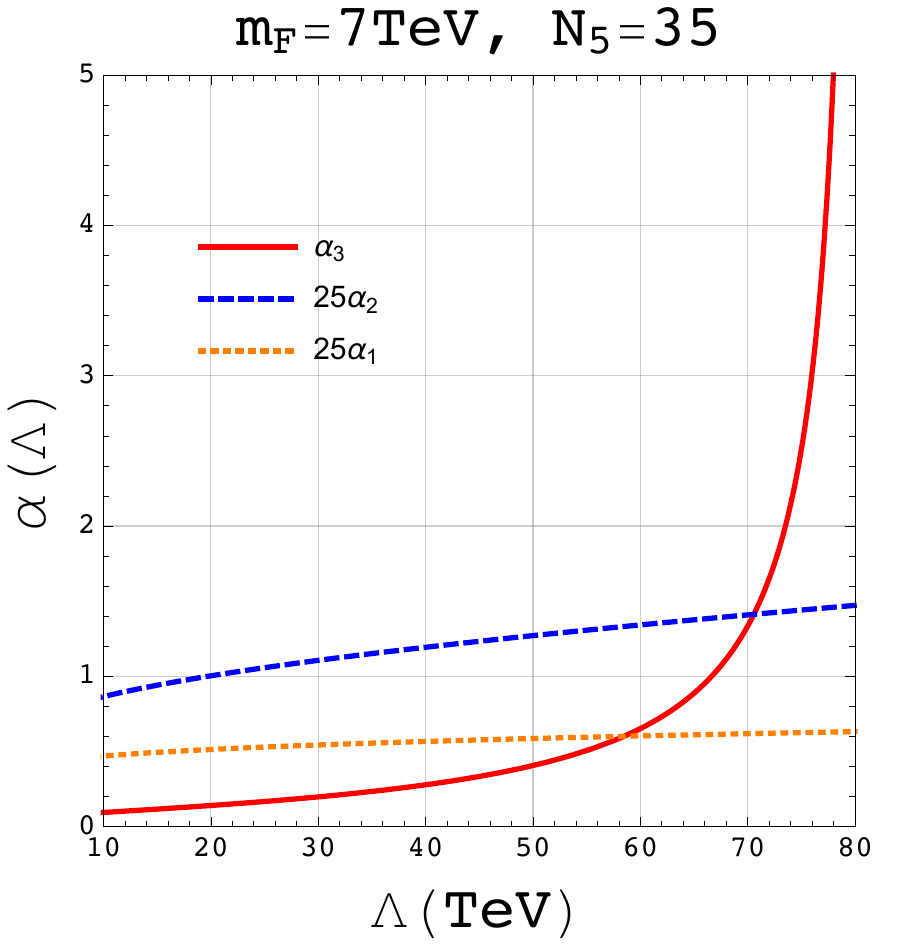} \\
\end{tabular}
\caption{{\sf RG running of the SM gauge couplings above $m_F$ for the two representative sets of values of \{$m_F$, $N_5$\} 
shown as black dots in Fig.~\ref{fig:OGM-param}, see text for more details. The values of the couplings at the scale $m_F$ is 
obtained using the SM evolution from $m_Z$ to $2 \TeV$ and the MSSM evoulution from $2\TeV$ to $m_F$.}
\label{fig:Running}}
\end{figure}

However, for such a large number of messenger fields, the gauge couplings lose asymptotic freedom. The one-loop running 
of the gauge couplings above the messenger fermion mass $m_F$ is shown in Fig.~\ref{fig:Running} for two sets of values 
of \{$m_F$, $N_5$\}, shown as black dots in Fig.~\ref{fig:OGM-param}. The  point \{$m_F$, $N_5$\} = \{$14\TeV$, $65$\} is chosen 
such that all the requirements namely, correct amount of the signal cross section and  heavy enough gaugino and squark masses 
are satisfied. 
It can be seen from the left panel of Fig.~\ref{fig:Running} that the $SU(3)$ gauge coupling in this case hits a one-loop Landau pole below 
$\sim 50 \TeV$.
The right panel of Fig.~\ref{fig:Running} shows the renormalisation group (RG) running for \{$m_F$, $N_5$\} = \{$7\TeV$, $35$\} i.e., 
when the constraint from the squark masses is relaxed. 
This is relevant for example, in models where the squark masses are generated at the tree level \cite{Nardecchia:2009ew,Nardecchia:2009nh}. 
However, even in this case, the required number of messenger pairs is $\gtrsim 35$ and the one-loop Landau pole is encountered below $\sim 80 \TeV$.

Before concluding this section, we would like to make two final comments:\\
i)Although we have presented our results for messengers 
transforming under $\{{\bf 5} + \overline{{\bf 5}}\}$ of $SU(5)$, our general conclusions hold for other representations also and even in the 
case when the possibility of doublet-triplet splitting is considered (this will be more clear in section~\ref{qqbar-production}). \\
ii)The formula in Eq.~\ref{eq:ci} is strictly valid only if the SUSY breaking VEV is small namely, $\kappa << 1$. For $\kappa \sim 1$, one has 
to compute the separate loop contributions from the messenger scalar with masses $m_\pm^2 = m_F^2( 1 \pm \kappa)$. This gives a correction factor 
$\sim \left(\dfrac{1 - 2/3 \, \kappa^2}{1-\kappa^2}\right)^2$ in the decay rates for the scalar $\phi$ ( here we have assumed $\lambda =1$ for simplicity). 
 This factor is only $\approx 2.5$ for $\kappa=0.8$ which we use for our analysis\footnote{The paper \cite{Baratella:2016daa} which appeared {\it after} 
 the first version of our paper  considered the very fine tuned possibility of $\kappa$ being extremely close to unity which may somewhat 
 mitigate the problem,  however, at the cost of  very large trilinear coupling between the sgoldstino and some of the light messenger scalars. 
 We do not consider this extremely fine-tuned possibility  in this paper.} and is absent for $a$.  Hence, this does not affect our numerical analysis.

\subsection{Possibility of larger $\lambda$}

It can be seen from Eq.~\ref{eq:ci} that, for a given gaugino mass, the $c_i$ coefficients (hence, diphoton signal cross section) can be 
increased  by increasing $\lambda$. However, one should first check the RG running of $\lambda$ in order to see the maximum 
value of $\lambda$ that is safe.

\begin{figure}[t]
\centering
\begin{tabular}{c c }
\includegraphics[scale=0.8]{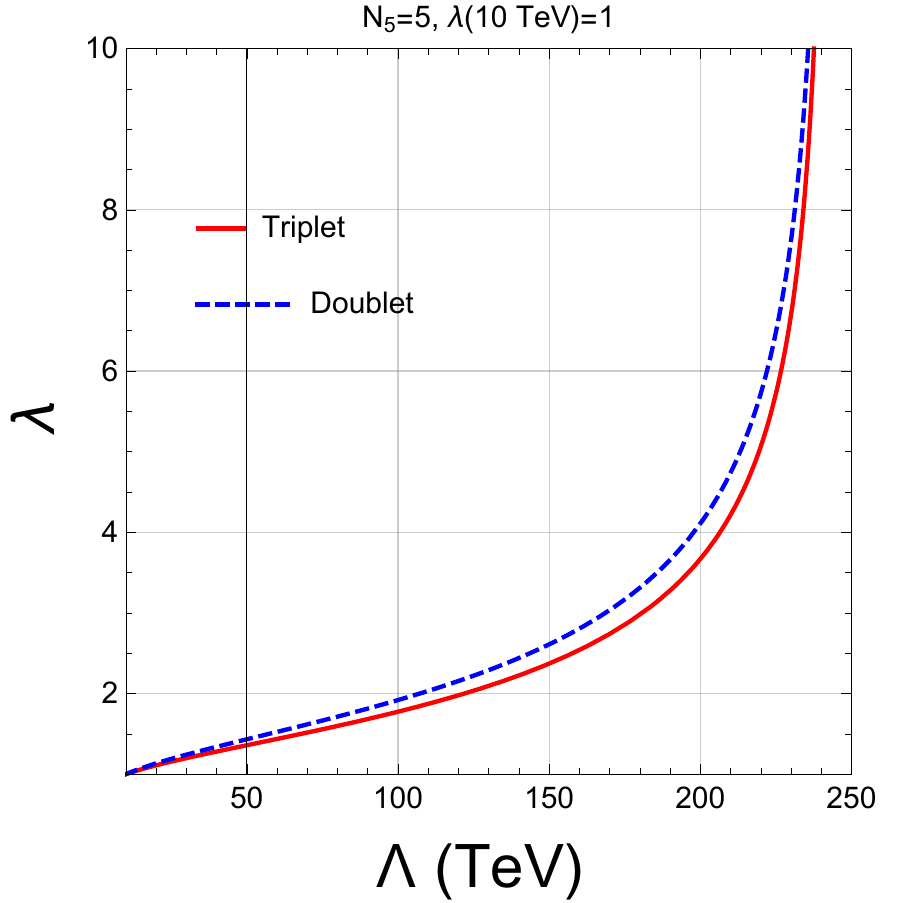} & ~~~~ \includegraphics[scale=0.8]{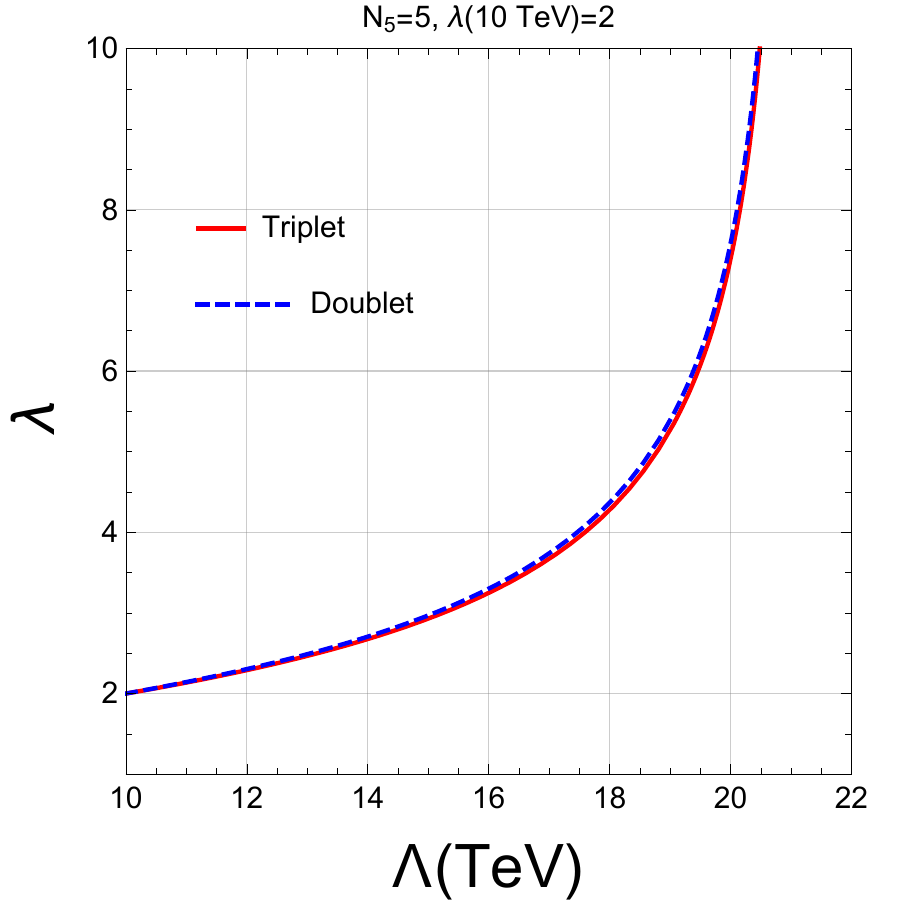} \\
\end{tabular}
\caption{{ \sf RG evolution of $\lambda$ for $N_5 =5$ and for two initial values of $\lambda$: $\lambda (10 \TeV) =1$ (left) and 
$\lambda (10 \TeV) =2$ (right).}}
\label{fig:Running_lambda}
\end{figure}

As the fundamental representation of $SU(5)$ can be decomposed into representations of $SU(3) \otimes SU(2) \otimes U(1)$ in the 
following way, 
\bea
{\bf 5} \to (3,1)_{-1/3} \oplus (1,2)_{1/2}
\eea
the superpotential can be rewritten as, 
\beqa
W= \lambda_{i}^{D^c}  X \Phi_i^{D^c} \tilde \Phi_i^{D^c} + \lambda_{i}^{L}  X 
\Phi_i^{L} \tilde \Phi_i^{L}
\eeqa
Note that, the notation $D^c$ and $L$ have been used just for notational convenience and they do not represent 
the MSSM fields.  
The beta functions of these couplings are given by
\beqa
\beta_{\lambda_{i}^{D^c}} &=& \lambda_{i}^{D^c} \left(\gamma(\Phi_i^{D^c}) + 
\gamma(\tilde \Phi_i^{D^c}) + \gamma(X)   \right) \, , \\*
\beta_{\lambda_{i}^{L}} &=& \lambda_{i}^{L} \left(\gamma(\Phi_i^{L}) + 
\gamma(\tilde \Phi_i^{L}) + \gamma(X)   \right),
\eeqa
where 
\beqa
 \gamma(\Phi_i^{D^c}) & =& \gamma(\tilde\Phi_i^{D^c})=  \frac{1}{4 \pi} (-\frac{2 \alpha 
_1}{15}+\alpha _{d \, i}-\frac{8 \alpha _3}{3}) \\
 \gamma(\Phi_i^{L}) & = & \gamma(\tilde \Phi_i^{L})=  \frac{1}{4 \pi } (-\frac{3 \alpha 
_1}{10}+\alpha_{l \, i}-\frac{3 \alpha _2}{2})\\
 \gamma(X) & =& \sum_i  \frac{1}{4 \pi }(3 \alpha _{d \, i}+2 \alpha_{l \, i})
\eeqa
We have used the notation, $\alpha _{d \, i} \equiv \dfrac{(\lambda_{i}^{D^c})^2}{4\pi}$ and $\alpha _{l \, i} 
\equiv \dfrac{(\lambda_{i}^{L})^2}{4\pi}$.

Hence, the RG equations for the $\lambda$ couplings are, 
\beqa
\frac{d\, \lambda_{i}^{D^c} }{dt} &=& \frac{1}{16\pi^2} \lambda_{i}^{D^c} \Big[ 
 (3\,N+2) \left(\lambda_{i}^{D^c}\right)^2 - \frac{16}{3} g_3^2 -\frac{4}{15} 
g_1^2 + 2 N (\lambda_{i}^{L})^2 \Big] \, ,\\
\frac{d\, \lambda_{i}^{L} }{dt} &=& \frac{1}{16\pi^2} \lambda_{i}^{L} \Big[  
(2\,N+2) \left(\lambda_{i}^{L}\right)^2 - 3 g_2^2 -\frac{3}{5} 
g_1^2 + 3 N (\lambda_{i}^{D^c})^2 \Big],
\eeqa

In Fig.~\ref{fig:Running_lambda} we show the running of these $\lambda$ couplings for five pairs of \{$\bf{5} + \bf{\bar{5}}$\} 
messengers and for two initial values of $\lambda$ at the scale $10 \TeV$, $\lambda (10 \TeV) =1 \, \, \text{and} \, \, 2$. It can be 
seen from the right panel of Fig.~\ref{fig:Running_lambda} that even for $\lambda (10 \TeV) =2$, it grows very fast and hits a one-loop 
Landau-pole below $\sim 25 \TeV$. Needless to say, the situation gets worse if a larger number of messenger pairs is considered. 
Hence, we conclude from this analysis that values of $\lambda$ much larger than unity at the messenger scale is not a possibility.

\subsection{Estimate of the mass of ${\cal S}$}
\label{sgoldstino-mass-1}
It was shown in \cite{Komargodski:2009jf} that in renormalizable Wess-Zumino models with canonical  K{\"a}hler potential, 
the existence of a massless fermion implies that the complex scalar  in the same chiral multiplet remains massless at the tree level 
even if SUSY is spontaneously broken. As the fermion component of $X$ is the goldstino in our case (which is exactly massless 
even at loop level), the scalar component of $X$, the sgoldstino will be massless at the tree level.  However, in general, the 
sgoldstino is expected to acquire non-zero mass when loop corrections are included. 

In our scenario, the sgoldstino mass gets contribution from the loops of messenger fields (apart from possible contributions from the 
hidden sector). The messenger contribution is computed in appendix~\ref{sgoldstino-mass}. The final result is given by 
(for $N_5$ pairs of ${\bf 5} + {\bf \bar{5}}$ of $SU(5)$),   
\bea
\Pi(p^2=0) &=&   - \left(\frac{\lambda}{g_3^2}\right)^2 \, \left(4 \pi \, \sqrt{\frac{5}{N_5}} F(x)\right)^2 \, m_{\tilde g}^2 
\eea

Hence, the potential for the sgoldstino gets a one-loop negative quadratic contribution from the messenger fields and this contribution 
is considerably larger in magnitude than the squared gluino mass\footnote{Note that, models with non-polynomial superpotential can give 
rise to tree level sgoldstino mass. We compute the sgoldstino mass in one such model \cite{Caracciolo:2012de} in 
appendix~\ref{tree-sgoldstino}, however, again it turns out to be in general much larger than the gluino mass.}. 
This means that a large contribution from the hidden sector 
is required to stabilise the sgoldstino potential and somehow generate a small mass $\sim 750 \GeV$ for the sgoldstino. 

At this point, we would like to remind the readers that, in our discussions till now, we have completely ignored specifying the details of 
the hidden sector  and how SUSY is broken there. We just assumed that the chiral superfield $X$ gets a SUSY breaking 
$F$-term VEV from the dynamics of the hidden sector without specifying the hidden sector at all. However, in order to understand 
whether a light sgoldstino can be obtained without too much tuning, we are now forced to consider the hidden sector as part of our 
model and think about the problem in its entirety.  We postpone any further investigation of this issue to section~\ref{way-out}.

\section{Extra Ordinary Gauge Mediation}
\label{EOGM}

We have seen in the previous section that the OGM framework needs a very large number of messengers in order to explain the 
diphoton signal and avoid the strong constraints on the gluino and squark masses from LHC. We have also seen that such a large 
number of messengers renders the theory non-perturbative at scales as low as $\sim 50 \TeV$, much below the GUT scale.

In this section we will consider a generalisation of the OGM framework namely, the Extra Ordinary Gauge Mediation (EOGM) 
where the OGM Lagrangian (Eq.~\ref{OGM-Lag}) is  supplemented with vector-like mass terms for the chiral superfiels  $\tilde \Phi_i$ 
and $\Phi_j$ \cite{Cheung:2007es}.
Hence, we now have the EOGM superpotential  
\begin{eqnarray}
\label{EOGM-Lag}
W_{{\rm EOGM}}= \left(\lambda_{ij} X  + m_{ij}  \right)\tilde \Phi_i \Phi_j   \, ,
\end{eqnarray}
where,  $\lambda_{ij}$ and $m_{ij}$ are arbitrary complex matrices. 
As in the OGM scenario, the auxiliary field of $X$ is assumed to get a VEV to break SUSY spontaneously. 
The fermion components of the messenger fields have the Dirac mass matrix, 
\bea
{m_F} = \lambda_{ij} \svev  + m_{ij} \, .
\eea
Without loss of generality, one can always go to the basis of $\tilde \Phi$  and $\Phi$ (by independent unitary rotations on them 
that do not affect their K{\"a}hler potential)  where $m_F$ is diagonal with real  eigenvalues $(m_F)_i$. Hence, from 
now on we will assume that the matrix $m_F$ is diagonal and the matrices $\lambda_{ij}$ and $m_{ij}$ are defined in the basis 
where $m_F$ is diagonal. The scalar mass-squared matrix in this basis can now be written as,  
\begin{eqnarray}
{\tilde m}^2 = \left(\begin{array}{cc}
{m_F}^2 & - \lambda \fvev \\
- \lambda \fvev & {m_F}^2
\end{array}
 \right) \, .
\end{eqnarray}
We will assume the matrix $\lambda$ to be real and symmetric in order to impose invariance under $CP$ and messenger parity 
(i.e., $\Phi_i \to \tilde \Phi_i$ in the basis where $m_F$ is diagonal) in the messenger sector \cite{Dvali:1996cu,Dimopoulos:1996ig}.

The matrix ${\tilde m}^2$ can be block diagonalised by a suitable change of basis of the scalar fields, the block diagonalised matrix being, 
\begin{eqnarray}
{\cal M}^2 = \left(\begin{array}{cc}
m_+^2 & 0 \\
0 & m_-^2
\end{array}
 \right),
\end{eqnarray}
where $m_\pm^2 = m_F^2 \pm \lambda \fvev$. 
Now assuming that the matrices $m_\pm^2$ are diagonalised by the unitary matrices $U_{\pm}$, the gaugino masses can be 
written as \cite{Marques:2009yu},
\begin{eqnarray}
m_a = \frac{\alpha_a}{4 \pi} \sum_\pm \sum_{i,j=1}^N (\pm)
(U_\pm^\dagger)_{ij} (U_\pm)_{ji} \, m_j \frac{m^2_{\pm i}{\rm Log}(m^2_{\pm i}/m_j^2)}{
m^2_{\pm i}-m_j^2}.
\end{eqnarray}

Let us now consider only one pair of messengers to simplify the discussion. In this case the expressions of the gaugino masses and 
couplings $c_a$ take the same form as the OGM case, 
\begin{eqnarray}
m_a &=& \frac{\alpha_a}{4 \pi} \, \kappa \, m_F \, g(\kappa)  \\
\frac{c_a}{M_a} &=& \frac{\alpha_a}{4 \pi} \, \frac{\lambda}{m_F} \, g(\kappa) 
\eea
the only difference being in the definition of $m_F$ which now has the form,
\bea
m_F = \lambda \svev + m \, .
\eea

Hence, for fixed values of the messenger fermion masses, the situation is exactly the same as OGM. 
In the presence of many pair of messengers, if $[m_F,\lambda] =0$ then the matrix $\lambda$ can be diagonalised simultaneously 
with $m_F$ and hence, the situation is again exactly  the same as OGM with many messenger fields. 
In the case when $[m_F,\lambda] \neq 0$, in general, one has to analyse the situation numerically. 
Analytic results are known even in this case for $\lambda \fvev << m_F^2$\cite{Cheung:2007es,Dumitrescu:2010ha}:
\begin{itemize}
\item The $R$ charge for the field $X$, $R(X) \neq 0$:  In this  case the expression of the gaugino mass 
can be written as, 
\bea
m_a = \frac{\alpha_a}{4 \pi} \, n_{{\rm eff}} \, \frac{\fvev}{\svev} 
\eea
where, 
\bea
n_{{\rm eff}} = \frac{1}{R(X)}\sum_i\left(2 - R(\Phi_i) - R(\tilde \Phi_i) \right) \, .
\eea
As $n_{{\rm eff}}$ is less than the total number of messengers, the gaugino mass in this case is always 
less than that in the OGM case.
\item $R(X)=0$, even in this case the expression of the gaugino mass simplifies to,
\bea
m_a = \frac{\alpha_a}{4 \pi} \, \fvev \sum_i \frac{\lambda_{ii}}{m_F^i},
\eea
If ${\rm \bf min}(m_F^i)= m$, then 
\bea
m_a \leq \frac{\alpha_a}{4 \pi} \, \frac{\fvev}{m} {\rm Tr}\lambda \, .
\eea
Hence, the situation is again the same as the OGM case. 
\end{itemize}

We have checked numerically that the situation does not improve for the case when $\lambda \fvev \sim m_F^2$.

\section{Way out?}
\label{way-out}

We have seen in the previous sections that an sgoldstino explanation of the diphoton excess faces two major issues: i) the gaugino masses, and in particular 
the gluino mass, turn out to be rather low unless a very large number of messenger fields is considered; ii) the messenger particles yield a large negative 
one loop contribution to the sgoldstino  potential.  In this section, our goal is to look for potential solutions of the above problems. 

\subsection{$D$-term contribution to the gaugino mass}

We have only considered $F$-term contribution to the gaugino mass in the previous sections. We will now assume that the messenger fields are also charged 
under some new $U(1)$ gauge group. The $\Phi$ fields have charge $+1$ and the $\tilde{\Phi}$ fields carry a charge $-1$ under this new $U(1)$. 
The relevant part of the Lagrangian is given by, 
\bea
{\cal L} \subset \int d^4\theta \, \left(\Phi^\dagger_i e^{gV} \Phi_i + \tilde{\Phi}^\dagger_i e^{-gV} \tilde{\Phi}_i\right)
+ \int d^2\theta \, \left(\lambda_{ij} X  + m_{ij}  \right)\tilde \Phi_i \Phi_j  + {\rm h.c.} \, .
\eea

The $F$-term of the chiral superfield $X$ and the $D$-term of the vector superfield $V$ are assumed to have VEVs $\fvev$ and $\langle D \rangle$ 
respectively\footnote{Note that the existence of non-zero $\langle D \rangle$ breaks the messenger parity spontaneously.}.  
However, since the above Lagrangian possesses an $U(1)$ R-symmetry, the 
charges being $R(\Phi) = 1, R(\tilde \Phi) = 1, R(X) = 0$ and  $R(V) =0$,  it follows that the $F$-term and the $D$-term have the $R$-charges 
$R(F) = 2$ and $R(D) =0$.  Hence, $\fvev \neq 0$ breaks $R$-symmetry spontaneously, while $\langle D \rangle \neq 0$ 
does not. It is then clear that the gaugino masses must be associated with non-zero $\fvev$.

As we discussed previously, the leading $F$-term contribution to the gaugino mass comes from the term 
\beqa
\label{D-term-gaugino-mass}
-\frac{1}{2} \, \frac{c_F}{\Lambda} \, X W_A W^A \, .
\eeqa
As the gaugino mass is always associated with $\fvev$, the $D$-term contribution must always be suppressed by higher powers of $\Lambda$ 
and hence, subdominant compared to the leading $F$-term contribution. That there is no $D$-term contribution at the leading order in the 
$F$-term VEV can also be understood diagrammatically. 
It can be seen from Fig.~\ref{fig:Dterm-gaugino} that, in order to join the scalar lines, one needs a 
term $\phi_1 \phi_2$ in the Lagrangian (refer to appendix~\ref{sgoldstino-mass} for the notation) which does not arise 
from the $D$-term. 

\begin{figure}[t]
\centering
\begin{tabular}{c}
\includegraphics[scale=0.65]{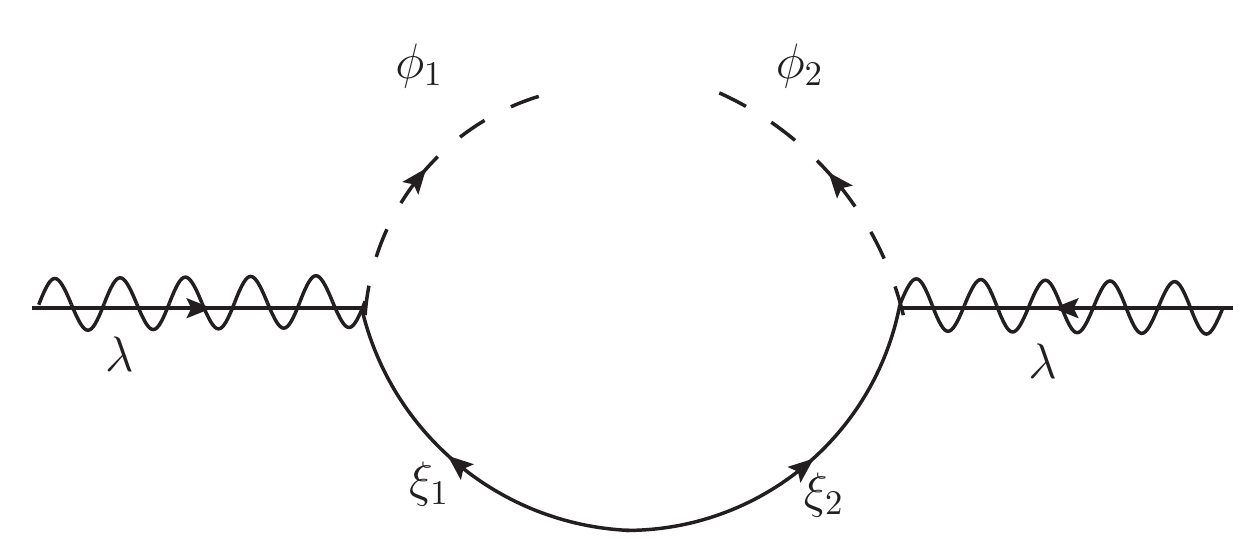} 
\end{tabular}
\caption{{ \sf Diagram showing that a $D$-term does not contribute to the the Majorana gaugino mass at the leading order in the 
$F$-term VEV.}}
\label{fig:Dterm-gaugino}
\end{figure}

In models with explicitly broken $R$-symmetry, the lowest dimensional operators that can give rise to the gaugino mass should be, 
\beqa
\label{DTC}
 -\frac{1}{2} \frac{c_D}{\Lambda_D^3} \, \widetilde{W} \widetilde{W} W_A W^A
\eeqa
which generates a contribution, 
\beqa
m_{\lambda} = c_D \frac{\langle D_{\widetilde W}\rangle^2}{\Lambda_D^3} \,, 
\eeqa 
which is subleading compared to \eqref{D-term-gaugino-mass}.
The chiral superfield $\widetilde W$ belongs to the hidden sector and corresponds to either an abelian or a non-abelian 
gauge group. Note that, as mentioned before, the term in \eqref{DTC} breaks $R$-symmetry explicitly. We thus conclude that 
$D$-term contribution can not enhance the gaugino mass considerably. 

We would like to comment in passing on the problem of vanishing leading order (in SUSY breaking $F$ term VEV) gaugino 
masses in models of direct gauge mediation \cite{Poppitz:1996fw,ArkaniHamed:1997jv} and semi-direct gauge mediation 
\cite{Seiberg:2008qj}, regardless of how the R-symmetry is broken.
The authors of \cite{Komargodski:2009jf} proved this in generalised renormalizable O'Raifeartaigh models assuming a 
locally stable pseudomoduli space. This problem can be avoided with non-polynomial superpotential which naturally appears 
in many models of dynamical/non-perturbative SUSY breaking (DSB) \cite{Affleck:1984xz,Peskin:1997qi,Poppitz:1998vd}. 
Hence, the gaugino mass to leading order in $\fvev$ that were considered in the previous sections should indeed be thought 
in the framework of DSB models.

\subsection{Metastable SUSY breaking}

Before going to the discussion of metastable SUSY breaking, it is worth reviewing briefly the relation between $R$-symmetry and 
spontaneous SUSY breaking.

Consider a generic model of gauge mediated supersymmetry breaking in which a 
{\it Hidden} sector (HS) consisting of the superfields $(Y_a,X)$ breaks supersymmetry and then {\it messenger} fields 
$(\Phi_i, \tilde{\Phi}_i)$ communicate the supersymmetry breaking to the {\it visible} MSSM sector via loop effects. 
The hidden sector fields are neutral under the Standard Model gauge group but could have its own gauge dynamics while 
the messenger fields $(\Phi_i, \tilde{\Phi}_i)$ transform in a vector like representation of SM gauge group 
and could also be charged under the HS gauge group. 

Let us write the full superpotential of the theory as follows
\beq\begin{split}
W &=  W_{{\rm HS}}(\{Y_a\},X)+ W_{{\rm M}}(X,\Phi_i,\tilde{\Phi}_i) + W_{{\rm MSSM}} , \\
\text{with} \quad & W_{M}= \lambda_{ij} X \Phi_i \tilde{\Phi}_j + m_{ij} \Phi_i \tilde{\Phi}_j \, . \\
\end{split}\eeq
Here $W_{{\rm MSSM}}$ is the MSSM superpotential and $W_{{\rm HS}}$ is hidden sector superpotential which 
spontaneously breaks SUSY\footnote{Note that the $R$-parity conserving MSSM has three parameter worth of 
$R$-symmetries. However, $R$-symmetry has gauge anomalies in the MSSM.}.  

What can one say about the $R$-symmetry in $W_{HS}$? Note that, for generic superpotential without $R$-symmetry, 
Nelson and Seiberg showed that a supersymmetric vacuum always exists \cite{Nelson:1993nf}. In other words, 
$R$-symmetry is a necessary (but not sufficient) condition for spontaneous breaking of supersymmetry. 
However, unbroken $R$-symmetry forbids  (Majorana) masses for the gauginos. Thus, it must be broken 
spontaneously which, in turn, would lead to a massless $R$-axion that may be dangerous for 
phenomenology\footnote{R-symmetry may be broken by Gravity effects, thus giving mass to the R-axion \cite{Bagger:1994hh}}.

Another possibility is to break $R$-symmetry explicitly in hidden sector ($W_{HS}$). Now it is possible to 
write down models with no $R$-symmetry which break SUSY spontaneously but these 
models have a non-generic superpotential in the sense that it doesn't allow all renormalisable 
terms allowed by symmetries. As superpotential couplings are protected from renormalisation 
and hence are not generated at loop levels, a non generic superpotential is technically natural. 
However, it is tuned and not satisfactory.

One scenario which avoids these problems is metastable supersymmetry breaking\cite{Intriligator:2007py}. 
It is based on the idea that though the true vacuum is supersymmetric, our universe lies in a metastable vacuum. In this 
picture, there is no need to keep $R$-symmetry but one does need to worry about decay rates from the metastable vacuum to the 
true vacuum and arrange for a long lived universe. 

As mentioned in the previous section, the problem of vanishing leading order (in SUSY breaking $F$-term) gaugino 
masses can be avoided in models of DSB. Hence, DSB in a metastable vacuum is an attractive phenomenological 
possibility. In fact, some of these models can potentially solve the problem mentioned in 
section~\ref{sgoldstino-mass-1} and give rise to a light sgoldstino \cite{Giveon:2008ne,Giveon:2009yu,Amariti:2008uz}. 
However, detailed exploration of these models is necessary to see whether they can indeed serve as natural models 
for a light sgoldstino and avoid the problems mentioned in section~\ref{OGM}.

\subsection{Quark anti-quark initiated production of the sgoldstino}
\label{qqbar-production}
In this section, we consider the possibility that the production cross section of sgoldstino has a significant contribution from 
quark anti-quark initial state. The coupling of the sgoldstino to the quark anti-quark pair can arise from the same effective Lagrangian 
that generates the trilinear $A$-terms namely,   
\begin{eqnarray}
\label{eff-Lag-2}
\mathcal{L}_{{\rm trilinear}} \subset 
\frac{A_u}{\fvev} \int d^2\theta X H_u Q U^c  +  \frac{A_d}{\fvev} \int d^2\theta X H_d Q D^c  \, \, + {\rm h.c.}
\end{eqnarray}
which generates following couplings for the sgoldstino,
\bea
\frac{v \sin\beta A_u }{\sqrt{2} \fvev} \, {\cal S} \, \bar{u} \, P_L \, u ~~, ~~ \frac{v \cos\beta A_d }{\sqrt{2} \fvev} \, {\cal S} \, \bar{d} \, P_L \, d \,. 
\eea

The decay rates $\Gamma(\phi \to \bar{u} \, u)$ and  $\Gamma( \phi \to \bar{d} \, d)$ can now be calculated from the above 
Lagrangian and read,
\bea
\Gamma(\phi \to \bar{u} \, u) &=&    \left[\frac{1}{2 m_\phi}\right] \left[\frac{1}{8 \pi}\right] \left[ \left(\frac{v \sin\beta A_u }{\sqrt{2}\fvev}\right)^2 
3m_\phi^2 \right] \, , \\ 
\Gamma(\phi \to \bar{d} \, d) &=&   \left[\frac{1}{2 m_\phi}\right] \left[\frac{1}{8 \pi}\right] \left[ \left(\frac{v \cos\beta A_d }{\sqrt{2}\fvev}\right)^2 
3m_\phi^2 \right] \,,
\eea
where we have neglected the quark masses. In this limit, the corresponding decay rates of $a$ have the same expressions with $m_\phi$ 
replaced by $m_a$.  

\begin{figure}[t]
\centering
\begin{tabular}{c c }
\includegraphics[scale=0.82]{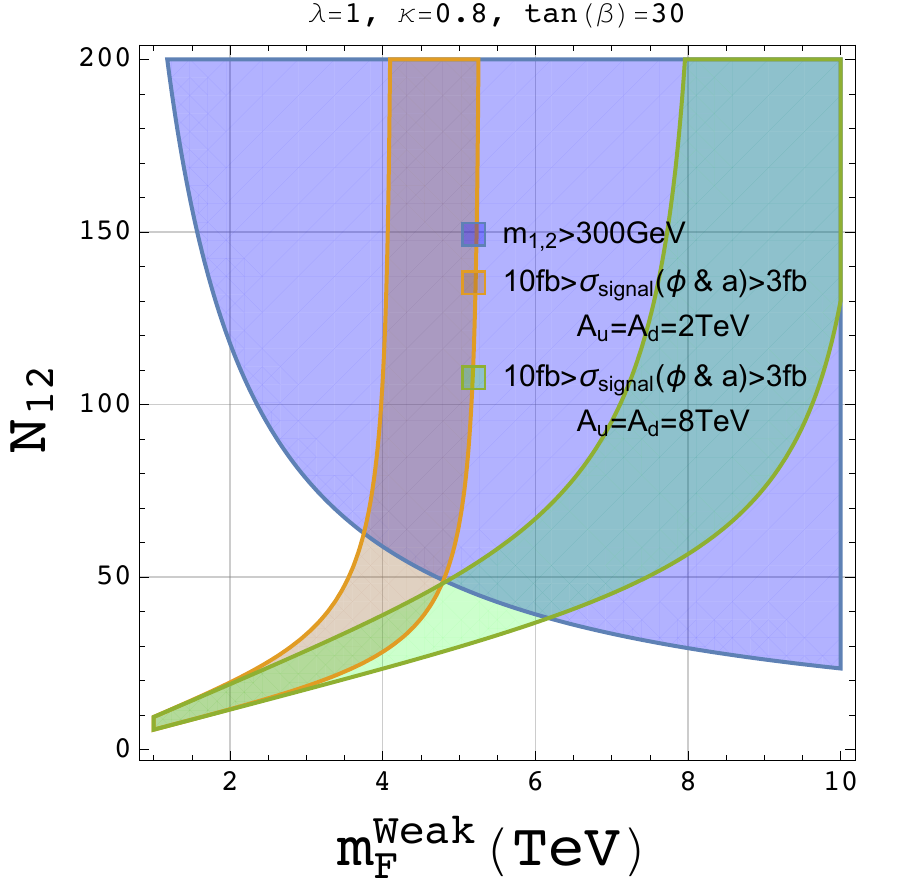} & ~~~\includegraphics[scale=0.82]{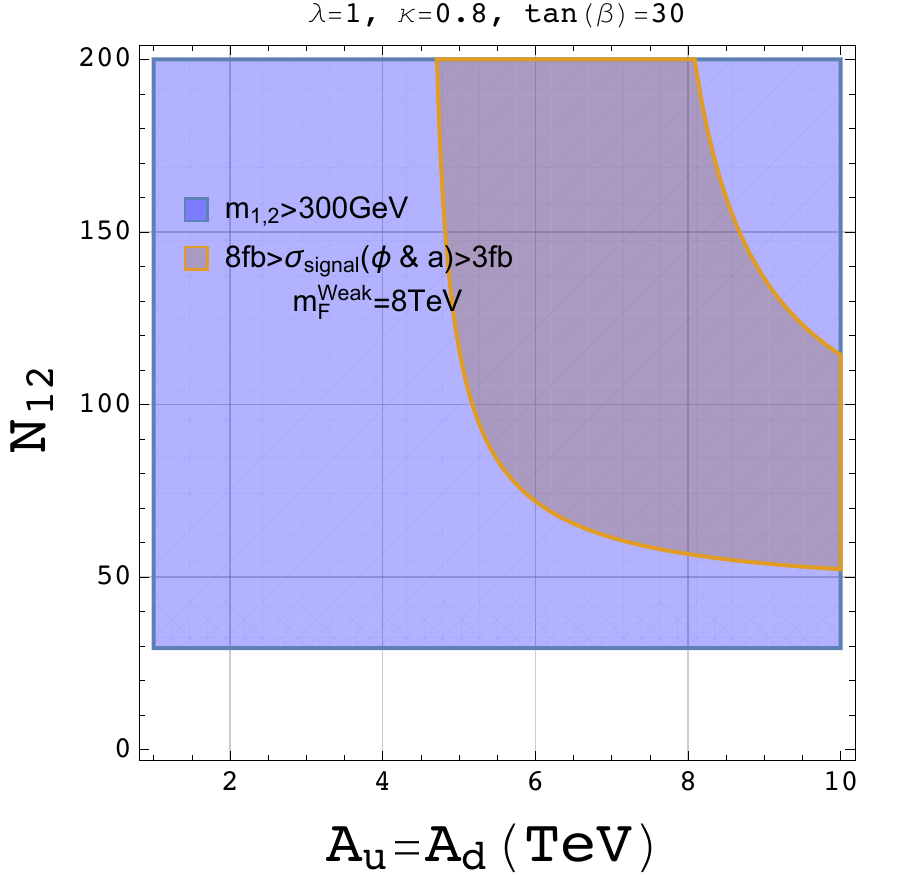}
\end{tabular}
\caption{{\sf Allowed region in the case of quark anti-quark initiated production of the sgoldstino. See text for more details.}}
\label{fig:qqbar}
\end{figure}

We now assume  that the production of sgoldstino is mostly by the $\bar{u}u$ and $\bar{d}d$ initial states so that a
large coupling to gluons is not necessary. We define the number of messengers with quantum numbers $(1,2)_{1/2}$ 
to be $N_{12}$. Their mass will be denoted by $m_F^{{\rm weak}}$.  
In the left panel of Fig.~\ref{fig:qqbar} we show the allowed region in the $N_{12}$ -- $m_F^{{\rm weak}}$ plane when two sets of values 
for $A_u$ and $A_d$ are chosen\footnote{In general, $A$-terms are generated at 1-loop level in the models of messenger matter interactions. 
Thus they are of same order of the gaugino masses. Larger $A$-terms can be obtained from model where $A$-terms are generated at the 
tree level \cite{Basirnia:2015vga}. These models have the advantage of being free from $A/m^2$ problem \cite{Craig:2012xp}.}. 
Similarly, in the right panel the allowed region in the $N_{12}$ -- $A_u/A_d$ plane is shown for $m_F^{{\rm weak}}=8 \TeV$. 
It can be seen that even for very large value of $A_u=A_d \sim 10 \TeV$
\footnote{Note that very large $A$-terms may give rise to electric charge and $SU(3)$ colour breaking minima in the potential 
\cite{Kusenko:1996jn,Chowdhury:2013dka}, thus we restrict them to $10 \TeV$ in our analysis.}, quite low masses for the electroweak 
messenger fields $m_F^{{\rm weak}} \lesssim 10 \TeV$ with a very large multiplicity $\gtrsim 50$ are necessary. Consequently, the $SU(2)$ and $U(1)$ 
couplings (i.e., $g_i^2/4\pi$) hit Landau poles typically below few hundred TeV. For example, for $m_F^{{\rm weak}} = 8 \TeV$ and  $N_{12} = 100$, the one 
loop Landau poles for $SU(2)$ and $U(1)$ appear around $50 \TeV$ and $200 \TeV$ respectively.

As the SUSY breaking $F$-term VEV $\fvev$ must be less than $(m_F^{{\rm weak}})^2$ in order to avoid tachyons in the messenger sector, it also turns out that 
a gluino mass of more than $1.5 \TeV$ again requires a very large number of $SU(3)$ messengers, exactly as in the OGM scenario discussed earlier.

However, one could consider a scenario where the $X$ superfields that couple to the $SU(3)$ messengers (denoted by $\Phi_3$ and $\tilde \Phi_3$ below) 
are different from the $X$ superfields that couple to the $SU(2)$ messengers (denoted by $\Phi_2$ and $\tilde \Phi_2$ below) so that, 
\begin{eqnarray}
\label{OGM-new}
W = (X_2 + m_2) \tilde \Phi_2 \Phi_2 + (X_3 + m_3) \tilde \Phi_3 \Phi_3  \, ,
\end{eqnarray}
The $X_2$ and $X_3$ superfields get VEVs given by,
\bea
\langle X_2 \rangle &=& \langle {\cal S}_2 \rangle + \theta \theta \langle F_2 \rangle \, ,\\
\langle X_3 \rangle &=& \langle {\cal S}_3 \rangle + \theta \theta \langle F_3 \rangle \, .
\eea
One can define two complex scalars that are linear combinations of ${\cal S}_2$ and ${\cal S}_3$, 
\bea
{\cal S}_h &=& \frac{F_2 {\cal S}_2 + F_3 {\cal S}_3}{\sqrt{F_2^2 + F_3^2}} \\
{\cal S}_l &=& \frac{-F_3 {\cal S}_2 + F_2 {\cal S}_3}{\sqrt{F_2^2 + F_3^2}}
\eea

\begin{figure}[t]
\centering
\begin{tabular}{c c }
\includegraphics[scale=0.82]{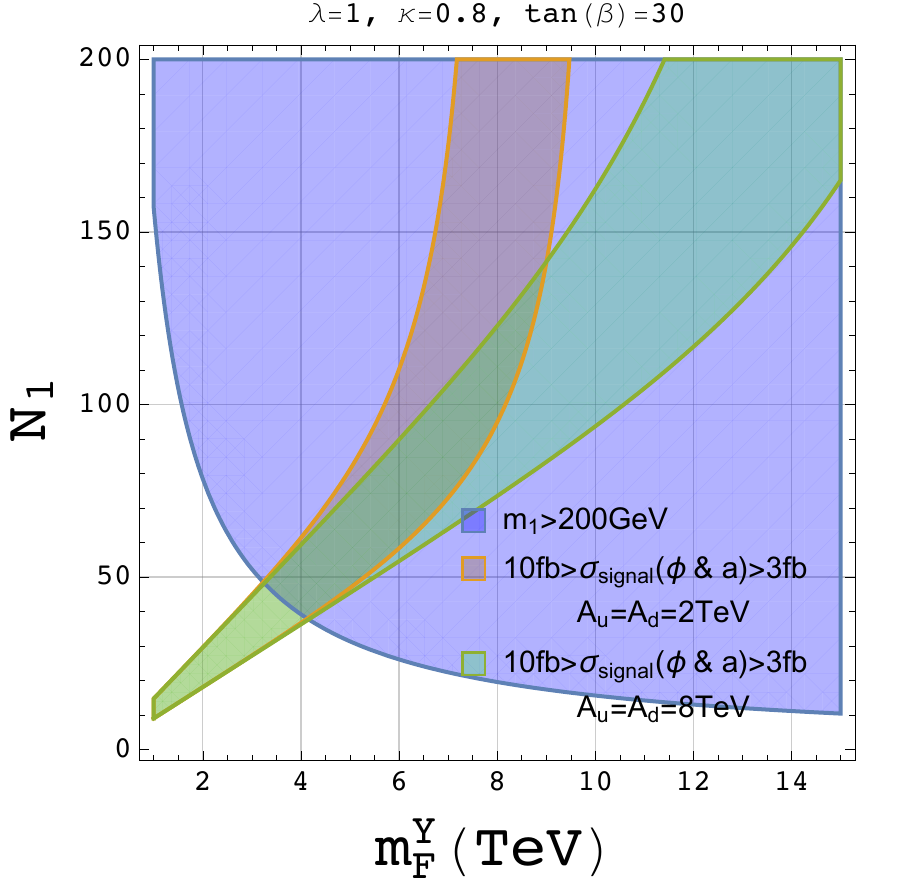} & ~~~\includegraphics[scale=0.82]{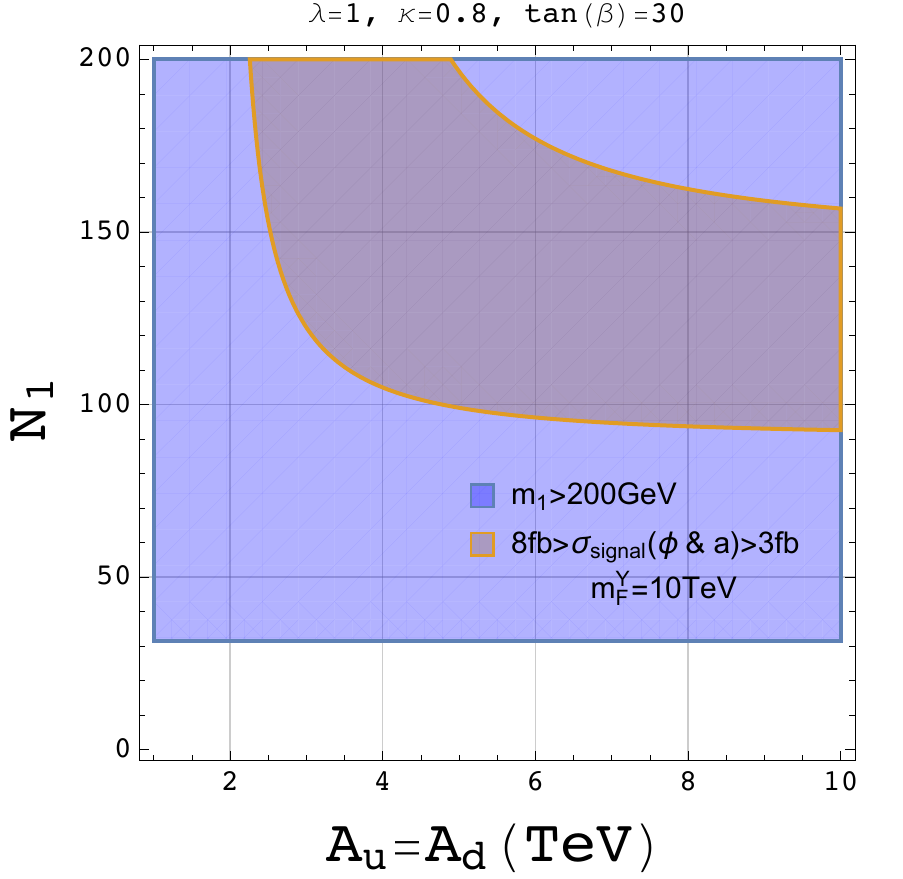}
\end{tabular}
\caption{{\sf Allowed region in the case of quark anti-quark initiated production of the sgoldstino with only light 
hypercharge messengers.}}
\label{fig:hypercharge}
\end{figure}

In the limit of $F_3 \gg F_2$, ${\cal S}_h \approx {\cal S}_3$ and ${\cal S}_l \approx {\cal S}_2$. If we now assume that the 
scalar ${\cal S}_l$ is actually the $750 \GeV$ resonance and the other scalar ${\cal S}_h$ is much heavier then the diphoton 
signal can be explained. Moreover, as $F_3$ is now assumed to be much large than $F_2$, large gluino mass can also be 
easily obtained. 

However, it should be mentioned that the scalar ${\cal S}_l$ is actually not the sgoldstino. It is actually ${\cal S}_h$ which appears 
in the goldstino multiplet, hence,  ${\cal S}_h$ should be identified as the sgoldstino. In this sense, we have not solved the original 
problem with sgoldstino being  the candidate for the $750 \GeV$ resonance.

Before concluding this section, we would also like to point out that one can also consider the extreme case when there are three 
different superfields $X_1$, $X_2$ and $X_3$ that couple to the $U(1)$, $SU(2)$ and $SU(3)$ messengers respectively. In this case, 
both the $SU(2)$ and $SU(3)$ messenger masses can be very high. In Fig.~\ref{fig:hypercharge} we show the number of 
$U(1)$ messengers ($N_1$) and their mass ($m_F^{{\rm Y}}$) required for the correct amount of signal and also mass of 
Bino more than $200 \GeV$. It can be seen that  for $m_F^{{\rm Y}} \sim 5 \TeV$ one needs $N_1 \gtrsim 50$.  The landau pole in 
the $U(1)$ gauge coupling only appears around $2000 \TeV$ in this case.

\section{Conclusion}
\label{conclusion}

In this paper we have carefully studied the possibility of an sgoldstino being a candidate for the signal of a possible new resonance with mass $\sim 750 \GeV$ 
recently reported by the ATLAS and CMS collaborations. We have found that the explanation of the signal is in tension with 
the lower bound on masses from direct searches of gauginos, in particular, the gluino. In order to achieve a large enough gluino mass, a very large number of messenger fields is required, which, in turn, renders the theory non-perturbative at a rather low scale of order few tens of TeV. 
Moreover, we find that the one-loop messenger contribution to the sgoldstino potential is negative and large in magnitude (larger than the gluino mass squared). 
Hence, a large positive contribution from the hidden sector is required to tune this away and get a small mass $\sim 750 \GeV$ for the sgoldstino.

While there exist examples of models with dynamically broken SUSY where a light sgoldstino can, in principle, be achieved, 
perhaps without large tuning, getting both the correct amount of signal cross-section and also large enough gluino and squark masses 
(without spoiling the calculability of the theory at a rather low scale) seems to be a stubborn problem.  
It would be interesting to find explicit models where these problems can be overcome in a satisfactory way. 
We postpone investigation in this direction to future studies.  

We have also considered the possibility of the resonance being produced by quark anti-quark initial state. While in this case the problem of Landau poles 
can be delayed beyond few thousand TeV, the scalar resonance can not be the sgoldstino.

\subsection*{Acknowledgements}
We thank Zohar Komargodski for many useful discussions and comments on this work. 
DG thanks Yevgeny Kats, Paride Paradisi, Fabrizio Senia and Fabio Zwirner for fruitful discussions. DB thanks Sabyasachi Chakraborty for several useful discussions. 

\begin{center}
{\Large \bf \underline{Appendix}}
\end{center}
\appendix

\section{Calculation of the partial decay widths}
\label{app-B}

In this appendix we will calculate the partial decay rate of $\phi$ and $a$ to two vector bosons.  

\subsection{$\phi \to \gamma \, \gamma$}
We start with the decay $\phi \to \gamma \, \gamma$ which arises from the following term in the Lagrangian, 
\begin{equation}
\mathcal{L} \subset \frac{1}{\Lambda}\phi F_{\mu\nu}F^{\mu\nu} \, .
\end{equation}

This yields the following Feynman rule,
\begin{multicols}{2}
\begin{align}
 \hspace*{12mm}\includegraphics[width=.3\textwidth]{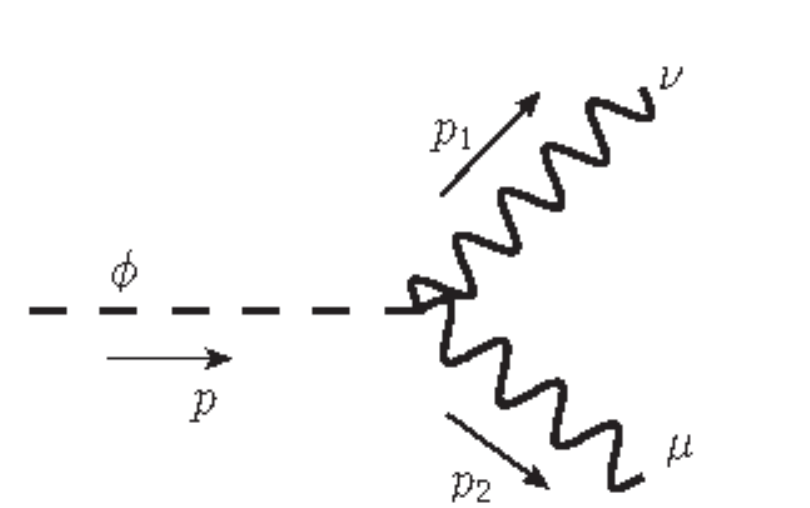} \nn 
\end{align}
\vspace*{5.5mm}
\begin{align}
\vspace*{1cm}
\hspace*{-15mm} [i]\,[2!] \, [- \frac{2}{\Lambda} \left(p_1 \cdot p_2 \, g_{\mu \nu} - p_{1\mu}p_{2\nu}\right)]
\label{phi-ga-ga-fr}
\end{align}
\end{multicols}

Thus, the matrix element is given by,
\begin{eqnarray}
i \mathcal{M} &=& - \frac{4}{\Lambda} \, i \,  \left(p_1 \cdot p_2 g_{\mu \nu} - p_{1\mu}p_{2\nu}\right) \epsilon^{*\nu}(p_1)\epsilon^{*\mu}(p_2) 
\end{eqnarray}

This gives, 
\begin{eqnarray}
\left|\mathcal{M}\right|^2 &=& \frac{16}{\Lambda^2} \left(p_1 \cdot p_2 g_{\mu \nu} - p_{1\mu}p_{2\nu}\right) \left(p_1 \cdot p_2 g_{\alpha \beta} - 
p_{1\alpha} p_{2\beta}\right) \epsilon^{*\nu}(p_1)\epsilon^{*\mu}(p_2)\epsilon^{\beta}(p_1)\epsilon^{\alpha}(p_2) \nn \\
\end{eqnarray}

Summing over the polarizations, i.e., 
\begin{equation}
\sum \epsilon^\mu(p) \epsilon^{*\nu}(p) = -g^{\mu\nu} \nn
\end{equation} 
we get,
\begin{eqnarray}
\left|\mathcal{M}\right|^2 &=& \frac{16}{\Lambda^2} \left(p_1 \cdot p_2 g_{\mu \nu} - p_{1\mu}p_{2\nu}\right) \left(p_1 \cdot p_2 g_{\alpha \beta} - 
p_{1\alpha} p_{2\beta}\right) g^{\alpha \mu} g^{\beta \nu} \nn \\
&=& \frac{16}{\Lambda^2} \left( p_1^2 \, p_2^2 + 4 (p_1 \cdot p_2)^2 - 2 (p_1 \cdot p_2)^2 \right) \nn \\
&=& \frac{32}{\Lambda^2} ( p_1 \cdot p_2)^2 \nn \\
&=& \frac{32}{\Lambda^2} \left(\frac{m_\phi^2}{2}\right)^2 \nn \\
&=& \frac{8 m_\phi^4}{\Lambda^2} 
\end{eqnarray}

Hence, 
\begin{eqnarray}
 \Gamma (\phi \to \gamma \, \gamma)
&=& \frac{1}{\Lambda^2} \left[\frac{1}{2 m_\phi}\right] \left[\frac{1}{8\pi}\right] \left[8 m_\phi^4\right]\left[\frac{1}{2}\right] \, .
\end{eqnarray}

The factor of 1/2 in the end is due to the presence of two identical particles in the final state.

\subsection{$a \to \gamma \, \gamma$}
The decay $a \to \gamma \, \gamma$ arises from the Lagrangian 
\begin{equation}
\mathcal{L} \subset   \frac{1}{\Lambda} a F_{\mu\nu}\tilde{F}^{\mu\nu} = \frac{1}{2 \Lambda} \, a \, F_{\mu\nu} F_{\alpha\beta} \, \varepsilon^{\mu\nu\alpha\beta}
\end{equation}

The Feynman rule for this vertex is given by 
\begin{multicols}{2}
\begin{align}
\hspace*{8mm}\includegraphics[width=.3\textwidth]{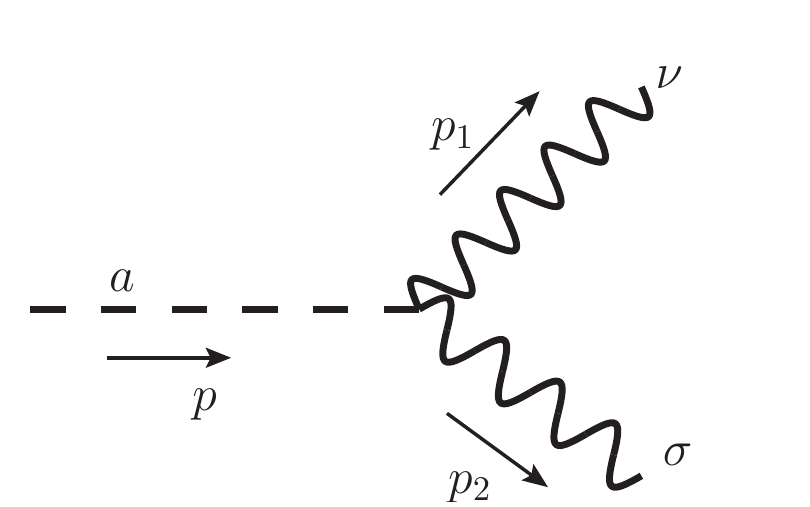} \nn 
\end{align}
\vspace*{5.5mm}
\begin{align}
\vspace*{1cm}
\hspace*{-15mm} [i]\,[2!] \, [- \frac{2}{\Lambda}\varepsilon^{\mu \nu \rho \sigma} p_{1\mu} p_{2\rho}]
\label{a-ga-ga-fr}
\end{align}
\end{multicols}

The matrix element and its square are given by,
\begin{eqnarray}
i \mathcal{M} &=& - \frac{4}{\Lambda}i  \, \varepsilon^{\mu \nu \rho \sigma} p_{1 \mu} p_{2 \rho} \epsilon^\ast_\nu (p_1)\epsilon^\ast_\sigma(p_2) \nn \\
\left|\mathcal{M}\right|^2 &=& \frac{16}{\Lambda^2} \, \varepsilon^{\mu \nu \rho \sigma} \varepsilon^{\alpha \beta \gamma \delta} p_{1 \mu}  p_{1 \alpha} p_{2 \rho}  p_{2 \gamma} \epsilon^\ast_\nu (p_1)\epsilon^\ast_\sigma(p_2) \epsilon_\beta (p_1)\epsilon_\delta(p_2)
\end{eqnarray}

Summing over the polarisations we get, 
\begin{eqnarray}
\sum \left|\mathcal{M}\right|^2 &=& \frac{16}{\Lambda^2} \,  \varepsilon^{\mu \nu \rho \sigma} \varepsilon^{\alpha \beta \gamma \delta} p_{1 \mu}  p_{1 \alpha} p_{2 \rho}  p_{2 \gamma} g_{\nu\beta} g_{\sigma \delta} \nn \\
&=& \frac{16}{\Lambda^2} \,  \varepsilon^{\mu \rho \sigma}_\beta  \varepsilon^{\beta \alpha \gamma}_\sigma p_{1 \mu}  p_{1 \alpha} p_{2 \rho}  p_{2 \gamma} g_{\nu\beta} g_{\sigma \delta} \nn \\
&=& \frac{32}{\Lambda^2} \, \left(- g^{\mu \alpha} g^{\rho \gamma} + g^{\mu \gamma} g^{\rho \alpha}\right) p_{1 \mu} p_{1 \alpha} p_{2 \rho} p_{2 \nu} \nn \\
&=& \frac{32}{\Lambda^2} \, (p_1\cdot p_2)^2 \nn \\
&=& \frac{32}{\Lambda^2} \left(\frac{m_a^2}{2}\right)^2 \nn \\
&=& \frac{8 m_a^4}{\Lambda^2} 
\end{eqnarray}

Hence, finally we get 
\begin{eqnarray}
 \Gamma (a \to \gamma \, \gamma)
&=& \frac{1}{\Lambda^2}\left[\frac{1}{2 m_a}\right] \left[\frac{1}{8\pi}\right] \left[8 m_a^4\right]\left[\frac{1}{2}\right]  \, .
\end{eqnarray}

\subsection{$\phi \to Z \, Z$ }

The relevant part of the Lagrangian is 
\begin{equation}
\mathcal{L} \subset \frac{1}{\Lambda} \phi Z_{\mu\nu}Z^{\mu\nu} \,. 
\end{equation}

The Feynman rule is same as the decay $\phi \to \gamma \gamma$ (Eq.~\ref{phi-ga-ga-fr}). 

The squared matrix element is given by, 
\begin{eqnarray}
\left|\mathcal{M}\right|^2 &=& \frac{16}{\Lambda^2} \left(p_1 \cdot p_2 g_{\mu \nu} - p_{1\mu}p_{2\nu}\right) \left(p_1 \cdot p_2 g_{\alpha \beta} - 
p_{1\alpha} p_{2\beta}\right) \epsilon^{*\nu}(p_1)\epsilon^{*\mu}(p_2)\epsilon^{\beta}(p_1)\epsilon^{\alpha}(p_2) \nn \\
\end{eqnarray}

Summation over the polarization vectors, 
\begin{equation}
\sum \epsilon^\mu(p) \epsilon^{*\nu}(p) = -g^{\mu\nu} + \frac{p^\mu p^\nu}{m_Z^2} \nn
\end{equation}
we get, 
\begin{eqnarray}
\left|\mathcal{M}\right|^2 &=& 16  \left(p_1 \cdot p_2 g_{\mu \nu} - p_{1\mu}p_{2\nu}\right) \left(p_1 \cdot p_2 g_{\alpha \beta} - p_{1\alpha} p_{2\beta}\right)
 \left(-g^{\mu \alpha} + \frac{p_2^\mu p_2^\alpha }{m_Z^2} \right)\left(-g^{\nu\beta} + \frac{p_1^\nu p_1^\beta}{m_Z^2} \right) \nn \\
&=&16  \left(p_1 \cdot p_2 g_{\mu \nu} - p_{1\mu}p_{2\nu}\right) \left(p_1 \cdot p_2 g_{\alpha \beta} - p_{1\alpha} p_{2\beta}\right)
 \left(-g^{\mu \alpha} \right)\left(-g^{\beta\nu} \right) \nn \\
&=& 16 \left(2 (p_1.p_2)^2 + p_1^2 p_2^2 \right) \nn \\
&=& 8 m_\phi^4 \left( 1 - 4 \frac{m_Z^2}{m_\phi^2} + 6 \frac{m_Z^4}{m_\phi^4} \right)
\end{eqnarray}

\begin{eqnarray}
 \Gamma (\phi \to Z \, Z)
&=& \left[\frac{1}{2 m_\phi}\right] \left[\frac{\lambda^{1/2}(1,m_Z^2/m_\phi^2,m_Z^2/m_\phi^2)}{8\pi}\right] 
|{\mathcal M}|^2\left[\frac{1}{2}\right] \\
\end{eqnarray}
where, 
\bea
\lambda(a,b,c) = a^2 + b^2 + c^2 - 2ab -2ac -2bc
\eea
\begin{eqnarray}
 \Gamma (\phi \to Z \, Z)
&=& \frac{1}{\Lambda^2} \left[\frac{1}{2 m_\phi}\right] \left[\frac{1}{8\pi} \left(1 - 4\frac{m_Z^2}{m_\phi^2}\right)^{1/2}\right] |{\mathcal M}|^2\left[\frac{1}{2}\right] \\
&=& \frac{1}{\Lambda^2} \left[\frac{1}{2 m_\phi}\right] \left[\frac{1}{8\pi} \left(1-\frac{4m_Z^2}{m_\phi^2}\right)^{1/2}\right] \nn \\
&&\hspace{2cm} \times \left[8 m_\phi^4 \left( 1 - 4 \frac{m_Z^2}{m_\phi^2} + 6 \frac{m_Z^4}{m_\phi^4} \right) \right] \left[\frac{1}{2}\right]
\end{eqnarray}

\subsection{$a \to Z \, Z$ }

The relevant part of the Lagrangian is 
\begin{equation}
\mathcal{L} \subset \frac{1}{\Lambda} a \, Z_{\mu\nu}\tilde{Z}^{\mu\nu} \,. 
\end{equation}

The Feynman rule is same as the decay $a \to \gamma \gamma$ (Eq.~\ref{a-ga-ga-fr}). 

The squared matrix element is given by, 
\begin{eqnarray}
\left|\mathcal{M}\right|^2 &=& \frac{16}{\Lambda^2} \varepsilon^{\mu \nu \rho \sigma} \varepsilon^{\alpha \beta \gamma \delta} 
p_{1\mu} p_{1\alpha} p_{2\rho} p_{2\gamma} \epsilon^{\ast}_\nu (p_1) \epsilon_{\beta}(p_1) \epsilon^\ast_\sigma(p_2) \epsilon_\delta(p_2)
\end{eqnarray}
Summing over the polarisations, we have, 
\begin{eqnarray}
\left|\mathcal{M}\right|^2 &=& \frac{16}{\Lambda^2} \varepsilon^{\mu \nu \rho \sigma} \varepsilon^{\alpha \beta \gamma \delta} 
p_{1\mu} p_{1\alpha} p_{2\rho} p_{2\gamma} \left(-g_{\nu \beta} + \frac{p_{1\nu} p_{1\beta}}{M_Z^2}\right)
\left(-g_{\sigma \delta} + \frac{p_{2\sigma} p_{2\delta}}{M_Z^2}\right) \\
&=& \frac{16}{\Lambda^2} \varepsilon^{\mu \nu \rho \sigma} \varepsilon^{\alpha \beta \gamma \delta} p_{1\mu} p_{1\alpha} 
p_{2\rho} p_{2\gamma} g_{\nu \beta} g_{\sigma \delta}
\end{eqnarray}
where we have used the fact that the second terms in each of the parenthesis vanish due to the anstisymmetry of the Levi-civita symbols. 
We thus have, 
\begin{eqnarray}
\left|\mathcal{M}\right|^2 &=& \frac{16}{\Lambda^2} \varepsilon^{\mu \rho \sigma}_\beta \varepsilon^{\beta \alpha \gamma}_\sigma 
p_{1\mu} p_{1\alpha} p_{2\rho} p_{2\gamma}
\end{eqnarray}
Using the relation, 
\begin{equation}
\varepsilon_\beta^{\mu \rho \sigma} \varepsilon^{\beta \alpha \gamma}_\sigma = 2 \left(-g^{\mu \alpha} g^{\rho \gamma} + g^{\mu \gamma}g^{\rho \alpha}\right)
\end{equation}
we get, 
\begin{eqnarray}
\left|\mathcal{M}\right|^2 &=& \frac{32}{\Lambda^2} \left(-g^{\mu \alpha} g^{\rho \gamma} + g^{\mu \gamma}g^{\rho \alpha}\right) 
p_{1\mu} p_{1\alpha} p_{2\rho} p_{2\gamma} \\
&=& \frac{32}{\Lambda^2} \left(-p_1^2 p_2^2 + (p_1 \cdot p_2)^2 \right) \\
&=& \frac{32}{\Lambda^2} \left(-m_Z^4 + \frac{(m_a^2 - 2 m_Z^2)^2}{4}\right) \\
&=& \frac{8 m_a^4}{\Lambda^2}  \left(1 - \frac{4 m_Z^2}{m_a^2}\right)
\end{eqnarray}
\begin{eqnarray}
 \Gamma (a \to Z \, Z)
&=& \left[\frac{1}{2 m_a}\right] \left[\frac{\lambda^{1/2}(1,m_Z^2/m_a^2,m_Z^2/m_a^2)}{8\pi}\right] |{\mathcal M}|^2\left[\frac{1}{2}\right] \\
&=& \frac{1}{\Lambda^2} \left[\frac{1}{2 m_a}\right] \left[\frac{1}{8\pi} \left(1 - 4\frac{m_Z^2}{m_a^2}\right)^{1/2}\right] |{\mathcal M}|^2\left[\frac{1}{2}\right] \\
&=& \frac{1}{\Lambda^2} \left[\frac{1}{2 m_a}\right] \left[\frac{1}{8\pi} \left(1-\frac{4m_Z^2}{m_a^2}\right)^{1/2}\right] \nn \\
&&\hspace{2cm} \times \left[8 m_a^4 \left( 1 - 4 \frac{m_Z^2}{m_a^2} \right) \right] \left[\frac{1}{2}\right]
\end{eqnarray}


\section{Calculation of $\mathcal A_{{\rm LHC \, \, energy}}^{ii}$}
\label{app-A}

In this appendix we will calculate the quantities $\mathcal A_{{\rm LHC \, \, energy}}^{ii}$ defined in section~\ref{excess} for two LHC energies $8 \TeV$ and 
$13 \TeV$, and for the initial states $\{gg\}$, $\{\bar{u}u\}$ and $\{\bar{d}d\}$. 

\subsection{Production by gluon fusion}

The partonic cross section for the process $g (p) \, g(k) \to \phi(q)$ is given by
\begin{eqnarray}
&&\hat{\sigma}(g (p) \, g(k) \to \phi(q)) \\ 
&=& \frac{1}{2^2} \frac{1}{8^2} \frac{1}{2 E_p  \, 2 E_k} \frac{1}{|v_p - v_k|} \int \frac{d^3q}{(2\pi)^3} \frac{1}{2E_q} 
\left|\mathcal{M}\right|^2 (2\pi)^4 \delta^{(4)}(p+k-q) \\
&\Rightarrow& \textnormal{using the identity} \int dq^0 \, \delta(q^2 - m_\phi^2) \,  \Theta(q^0) = \dfrac{1}{2 E_q}, \, \textnormal{we get} \\
&=&\frac{1}{2^2} \frac{1}{8^2} \frac{2 \pi }{2 E_p \, 2E_k} \frac{1}{|v_p - v_k|}   \int d^4q \, 
\delta(q^2 - m_\phi^2)  \, \Theta(q^0) \, \left|\mathcal{M}\right|^2 \, \delta^{(4)}(p+k-q) \\
&=& \frac{1}{2^2} \frac{1}{8^2} \frac{2 \pi }{2 E_p \, 2E_k} \frac{1}{|v_p - v_k|}  \left|\mathcal{M}\right|^2 \delta((p+k)^2 - m_\phi^2) \\
&=& \frac{1}{2^2} \frac{1}{8^2} \frac{2 \pi }{x_1 x_2 S } \frac{1}{2}   \left|\mathcal{M}\right|^2 \delta(x_1 x_2 S - m_\phi^2) \\
&=& \frac{\pi}{256} \frac{1}{x_1 x_2 S }  \left|\mathcal{M}\right|^2 \delta(x_1 x_2 S - m_\phi^2)
\end{eqnarray}

where the following definitions have been used,
$$
p=x_1 P_1, ~ k = x_2 P_2, ~ P_1=\dfrac{\sqrt{S}}{2}(1,0,0,1) ~{\rm and} ~ P_2=\dfrac{\sqrt{S}}{2}(1,0,0,-1).
$$
Here, $P_1$ and $P_2$ are the 4-momenta of the two protons and $\sqrt{S}$ is their centre-of-mass energy.

We now proceed to compute the hadronic cross section which is given by
\begin{eqnarray}
\sigma_{_{\sqrt{S}}} &=&  \int_0^1 dx_1 \int_0^1 dx_2  \, f_{g/p}(x_1) f_{g/p}(x_2) \, \hat{\sigma}(x_1,x_2) \\
&\Rightarrow& \textnormal{using the change of variables} \,  \{x_1, x_2\} \to \{x=x_1,z=x_1 x_2\} , \textnormal{we get} \nonumber\\
 &=& \int_0^1 \frac{dx}{x} \int_0^x dz  \, f_{g/p}(x) f_{g/p}(z/x) \, \hat{\sigma}(z) \\
 &=& \int_0^1 \frac{dx}{x} \int_0^x dz  \, f_{g/p}(x) f_{g/p}(z/x)  \times \frac{\pi}{256} \frac{1}{z \, S }  \left|\mathcal{M}\right|^2 \delta(z S - m_\phi^2) 
 \eea
 
 We now use the expression  for $\Gamma_{\phi \to g \, g}$ (following  appendix~\ref{app-B}),  
 \begin{eqnarray}
 \label{Gamma_gg}
 \Gamma (\phi \to g \, g)
&=&\left[\frac{1}{2 m_\phi}\right] \left[\frac{1}{8 \pi}\right] \left|\mathcal{M}\right|^2 \, \left[ \frac{1}{2}\right]\, ,
\end{eqnarray}
 to get 
 \bea
 \sigma_{_{\sqrt{S}}} &=&  \frac{\pi}{256} \frac{32 \, \pi \, m_\phi \, \Gamma_{\phi \to g \, g}}{S }   \int_0^1 \frac{dx}{x} \int_0^x dz  \, 
 f_{g/p}(x) f_{g/p}(z/x) \, \frac{1}{z}  \, \delta(z S - m_\phi^2) \\
 &=&  \frac{\pi}{256} \frac{32 \, \pi \, m_\phi \, \Gamma_{\phi \to g \, g}}{S^2 }   \int_0^1 \frac{dx}{x} \int_0^x dz  \, f_{g/p}(x) f_{g/p}(z/x) \, 
 \frac{1}{z}  \, \delta(z - \frac{m_\phi^2}{S}) \\
 &=&  \frac{\pi}{256} \frac{32 \, \pi \, m_\phi \, \Gamma_{\phi \to g \, g}}{S^2 }   \int_\frac{m_\phi^2}{S}^1 \frac{dx}{x}   \, f_{g/p}(x) f_{g/p}(m_\phi^2/Sx) \, 
\frac{S}{m_\phi^2}\\
&=&  \frac{\pi^2}{8} \frac{\Gamma_{\phi \to g \, g}}{m_\phi S }   \int_\frac{m_\phi^2}{S}^1 \frac{dx}{x}   \, f_{g/p}(x) f_{g/p}(m_\phi^2/Sx)
\end{eqnarray}

Hence, 
\bea
\mathcal A_{{\rm LHC \, \, energy}}^{gg} = \frac{\pi^2}{8} \frac{1}{m_\phi S }   \int_\frac{m_\phi^2}{S}^1 \frac{dx}{x}   \, f_{g/p}(x) f_{g/p}(m_\phi^2/Sx) \, 
\eea

Using the MSTW 2008 LO parton distribution functions (PDF) we get,  

\bea
\mathcal A_{13 \TeV}^{gg} &=& \frac{5.44 \pb}{\GeV} \\
\mathcal A_{8 \TeV}^{gg} &=& \frac{1.15 \pb}{\GeV} \, .
\eea

\subsection{Production by quarks}

The cross section of the process $\bar{q} \, q \to \phi$  can be calculated in the same way as above, except for the following changes, 
\begin{itemize}
\item The colour factor is different, so we must have $1/{3^2}$ instead of $1/{8^2}$ as in the case for gluons 
\item The symmetry factor (1/2) for identical particle used in Eq.~\eqref{Gamma_gg} no longer applies 
\item The PDF are different - we now have quark PDF instead of the gluon PDF.
\end{itemize}

Applying the above changes, we finally get,
\begin{eqnarray}
\sigma_{\sqrt{S}} 
&=& \frac{4 \pi^2}{9} \frac{\Gamma_{\phi \to q \bar{q}}}{m_\phi S} \int_{\frac{m_\phi^2}{S}}^1 \frac{dx}{x} \left(f_{q/p}(x)f_{\bar{q}/p}(m_\phi^2/{Sx}) + 
f_{\bar{q}/p}(x)f_{q/p}(m_\phi^2/{Sx})\right) \nn
\end{eqnarray}

Hence, 
\bea
\mathcal A_{{\rm LHC \, \, energy}}^{q \bar{q}} = \frac{4 \pi^2}{9} \frac{1}{m_\phi S} \int_{\frac{m_\phi^2}{S}}^1 \frac{dx}{x} \left(f_{q/p}(x)f_{\bar{q}/p}(m_\phi^2/{Sx}) + 
f_{\bar{q}/p}(x)f_{q/p}(m_\phi^2/{Sx})\right) \nn \, 
\eea

Using again the MSTW 2008 LO parton distribution functions (PDF) we get,  

\begin{multicols}{2}
\noindent
\begin{align}
{\mathcal A}_{13}^{\bar{u}u} \equiv {\mathcal A}|_{{\rm 13\,TeV\,LHC}}^{\bar{u}u} &=&  \dfrac{ 2.94 \pb}{\GeV}  \nn \\
{\mathcal A}_{13}^{\bar{d}d} \equiv {\mathcal A}|_{{\rm 13\,TeV\,LHC}}^{\bar{d}d} &=&  \dfrac{ 1.73 \pb}{\GeV} \nn \\
{\mathcal A}_{13}^{\bar{c}c} \equiv {\mathcal A}|_{{\rm 13\,TeV\,LHC}}^{\bar{c}c} &=&  \dfrac{ 0.11\pb}{\GeV} \nn \\
{\mathcal A}_{13}^{\bar{s}s} \equiv {\mathcal A}|_{{\rm 13\,TeV\,LHC}}^{\bar{s}s} &=&  \dfrac{ 0.21\pb}{\GeV} \nn \\
{\mathcal A}_{13}^{\bar{b}b} \equiv {\mathcal A}|_{{\rm 13\,TeV\,LHC}}^{\bar{b}b} &=&  \dfrac{ 0.05\pb}{\GeV} \nn
\end{align}
\begin{align}
{\mathcal A}_{8}^{\bar{u}u} \equiv {\mathcal A}|_{{\rm 8\,TeV\,LHC}}^{\bar{u}u} &=&  \dfrac{1.2 \pb}{\GeV} \\
{\mathcal A}_{8}^{\bar{d}d} \equiv {\mathcal A}|_{{\rm 8\,TeV\,LHC}}^{\bar{d}d} &=&  \dfrac{ 0.66 \pb}{\GeV} \\
{\mathcal A}_{8}^{\bar{c}c} \equiv {\mathcal A}|_{{\rm 8\,TeV\,LHC}}^{\bar{c}c} &=&  \dfrac{  0.03\pb}{\GeV} \\
{\mathcal A}_{8}^{\bar{s}s} \equiv {\mathcal A}|_{{\rm 8\,TeV\,LHC}}^{\bar{s}s} &=&  \dfrac{  0.05\pb}{\GeV} \\
{\mathcal A}_{8}^{\bar{b}b} \equiv {\mathcal A}|_{{\rm 8\,TeV\,LHC}}^{\bar{b}b} &=&  \dfrac{  0.01\pb}{\GeV}
\end{align}
\end{multicols}

\section{Calculation of the sgoldstino mass}
\label{sgoldstino-mass}
In this appendix, we want to compute the 1-loop contribution to the sgoldstino mass from the term, 
\bea
{\cal L} \subset  \int d^2 \theta \, \lambda X \Phi_1 \Phi_2  + \rm h.c.
\eea
We will ignore the gauge indices of $\Phi_1$ and $\Phi_2$ for the time being.  
The following notation will be used for the chiral superfields:
\begin{eqnarray}
X &=& {\cal S} + \sqrt{2} \, \theta \, \psi_x + \theta \theta \, F_x \\
\Phi_1 &=& \phi_1 + \sqrt{2} \, \theta \, \xi_1 + \theta \theta \, F_1 \\
\Phi_2 &=& \phi_2 + \sqrt{2} \, \theta \, \xi_2 + \theta \theta \, F_2
\end{eqnarray}

A Dirac fermion $\Psi$ is constructed out of the two Weyl fermions $\xi_1$ and $\xi_2$, 
\bea
\Psi = \left( 
\begin{array}{c}
\xi_{1\alpha} \\ 
\xi_{2}^{\dagger \dot \alpha}\end{array}
\right)
\eea
whose Dirac mass will be denoted by $m_\Psi = \lambda \langle {\cal S} \rangle$. The scalar mass eigenstates will be denoted 
by $\phi_+$ and $\phi_-$ with their mass squared 
given by $m_{\pm}^2 = m_\Psi^2 \pm \lambda \langle F_x \rangle$. 

\subsection{Diagrammatic calculation}

The relevant vertex factors are given by, 
\bea
{\cal S} \, \bar{\Psi} \, \Psi &:&  -\lambda P_L\\
{\cal S}^* {\cal S} \, \phi_+^* \phi_+ &:&  - \lambda^2 \\
{\cal S}^* {\cal S} \, \phi_-^* \phi_- &:&  - \lambda^2 \\
{\cal S} \, \phi_+^* \phi_+ &:&  -\lambda m_{\Psi} \\
{\cal S} \, \phi_-^* \phi_- &:& -\lambda m_{\Psi}
\eea

The Feynman rules can be obtained by multiplying the above vertex factors by $i$ and appropriate symmetry factors.

The relevant diagrams are, 
\begin{figure}[h]
\centering
\begin{tabular}{c c c}
\includegraphics[scale=0.55]{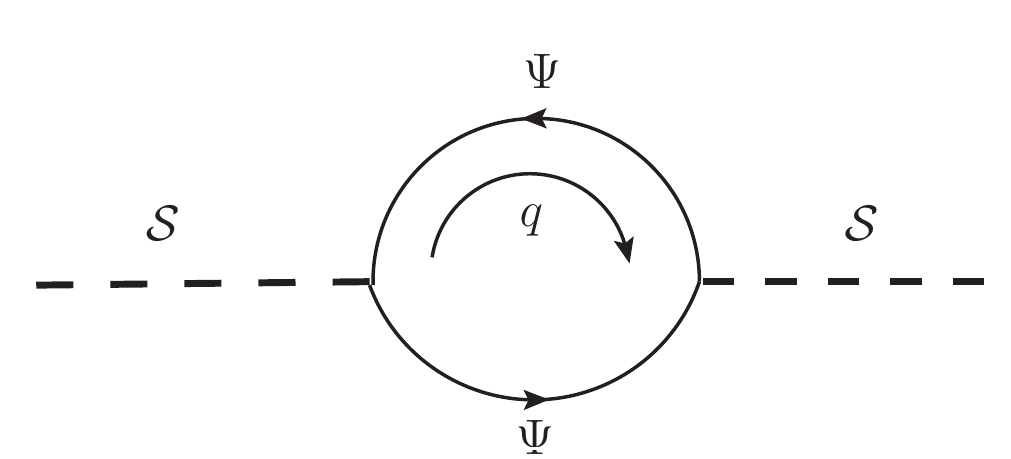} & 
\includegraphics[scale=0.6]{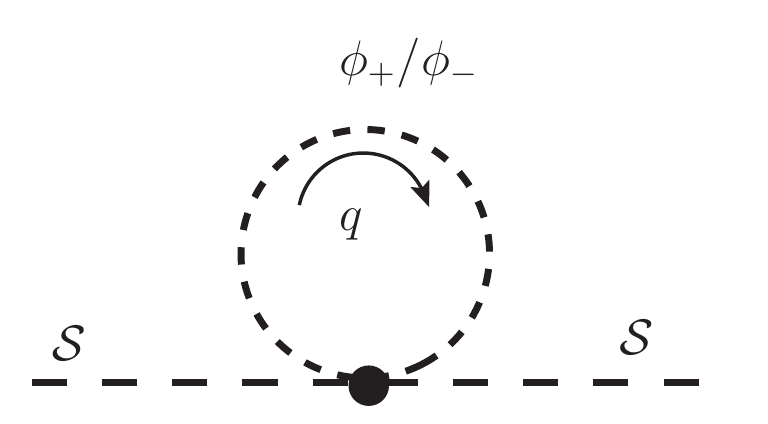}  & 
\includegraphics[scale=0.50]{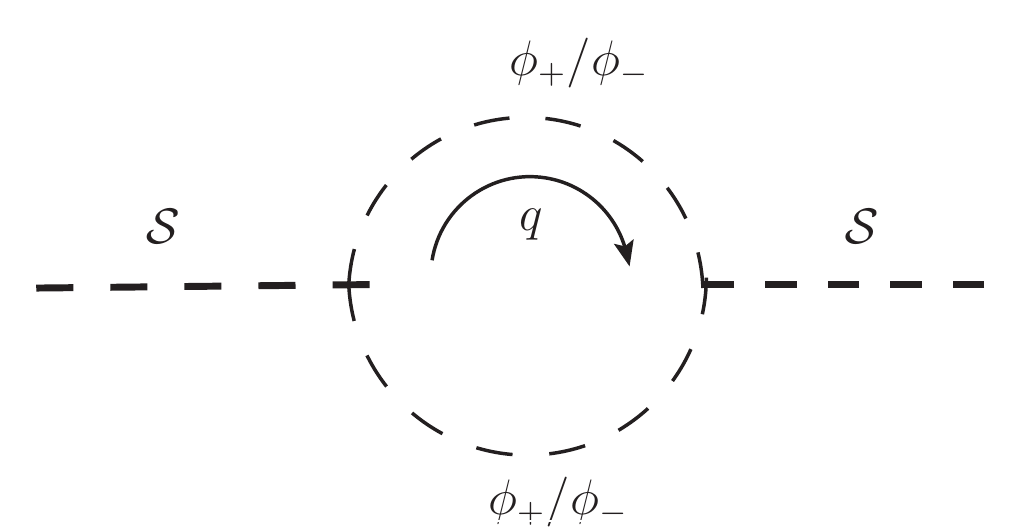}
\end{tabular}
\caption{{ \sf One loop contributions to the sgoldstino mass from the messengers.}}
\label{fig:sgoldstino-mass}
\end{figure}

We will now compute the diagrams one-by-one.  
\subsubsection*{Fermion loop} 
\begin{eqnarray}
-i \, \Pi(p^2=0) &=& - (- i\lambda)(- i\lambda) \int \frac{d^4q}{(2\pi)^4} {\rm Tr} \left[P_L\frac{i}{\not{\!q} - m_\Psi} P_R\frac{i}{\not{\!q} - m_\Psi}\right] \nonumber \\
&=& -2 \lambda^2 \int \frac{d^4 q}{(2\pi)^4} \frac{q^2}{\left(q^2 - m_\Psi^2\right)^2} \nn \\
&=& -2 \lambda^2 \int \frac{d^4q}{(2 \pi)^4} \left[\frac{1}{q^2 - m_\Psi^2} + \frac{m_\Psi^2}{\left(q^2 - m_\Psi^2\right)^2}\right]
\end{eqnarray}
\subsubsection*{First scalar loop} 
\begin{eqnarray}
-i \, \Pi(p^2=0) &=& (- i \, \lambda^2) \sum_{\phi=\phi_\pm} \int \frac{d^4q}{(2\pi)^4} \frac{i}{q^2 - m_\phi^2}  \nn \\
&=&  \lambda^2 \int \frac{d^4q}{(2\pi)^4} \left[\frac{1}{q^2 - m_+^2} + \frac{1}{q^2 - m_-^2}\right]
\end{eqnarray}

\subsubsection*{Second scalar loop} 
\begin{eqnarray}
-i \, \Pi(p^2=0) &=& (- i \lambda m_\Psi)(- i \lambda m_\Psi)  \sum_{\phi=\phi_\pm} \int \frac{d^4 q }{(2\pi)^4} \frac{i^2}{(q^2 - m_\phi^2)^2} \\
&=& \lambda^2 m_\Psi^2 \int \frac{d^4 q }{(2\pi)^4} \left[ \frac{1}{(q^2 - m_+^2)^2} + \frac{1}{(q^2 - m_-^2)^2}\right]
\end{eqnarray}
Note that the sum of the diagrams goes to zero in the limit of equal masses for the scalars and fermions, i.e. when SUSY is unbroken.

We need to evaluate integrals of two the forms: 
\begin{equation}
A_0(m) = \int \frac{d^D q}{(2\pi)^D} \frac{1}{q^2 - m^2}\ \ ; \ \ B_0(0,m,m) = \int \frac{d^D q}{(2\pi)^D} \frac{1}{\left(q^2 - m^2\right)^2}
\end{equation}
They are given by, 
\bea
A_0(m) &=& \frac{i}{16 \pi^2} \, m^2 \left[\frac{1}{\hat{\epsilon}} + 1 - {\rm Ln}\frac{m^2}{\mu^2}\right] \\
B_0(0,m,m) &=& \frac{A_0(m)}{m^2} - \frac{i}{16 \pi^2 }
\eea
where, 
\bea
\frac{1}{\hat{\epsilon}} = \frac{2}{4-D} - \gamma_E + {\rm Ln}(4\pi), \, \text{$\gamma_E$ being the Euler constant.}
\eea

Putting all loop contributions in order, we have 
\begin{eqnarray}
-i \, \Pi(p^2=0) &=& - 2 \lambda^2 \int \frac{d^4 q}{(2\pi)^4} \left[\frac{1}{q^2 - m_\Psi^2} - 
\frac{1}{2}\  \frac{1}{q^2 - m_+^2} - \frac{1}{2}\ \frac{1}{q^2 - m_-^2} \right.  \\ 
&&\ \qquad \left. + m_\Psi^2 \  \frac{1}{\left( q^2 - m_\Psi^2 \right)^2} -  \frac{m_\Psi^2}{2} \ \frac{1}{(q^2 - m_+^2)^2} - 
 \frac{m_\Psi^2}{2} \ \frac{1}{(q^2 - m_-^2)^2}\right] \nn \\
&=& - 2 \lambda^2 \frac{i}{16 \pi^2} \left[ m_\Psi^2 \left(\frac{1}{\hat{\epsilon}} + 1 - {\rm Ln}\frac{m_\Psi^2}{\mu^2}\right)
 - \frac{m_+^2}{2} \left(\frac{1}{\hat{\epsilon}} + 1 - {\rm Ln}\frac{m_+^2}{\mu^2}\right) \right. \nn \\
 && \left. ~~~~~~~~~~~~~~  - \frac{m_-^2}{2} \left(\frac{1}{\hat{\epsilon}} + 1 - {\rm Ln}\frac{m_-^2}{\mu^2}\right) 
 +  m_\Psi^2 \left(\frac{1}{\hat{\epsilon}} - {\rm Ln}\frac{m_\Psi^2}{\mu^2}\right) \right. \nn \\
 && \left. ~~~~~~~~~~~~~~ -  \frac{m_\Psi^2}{2} \left(\frac{1}{\hat{\epsilon}} - {\rm Ln}\frac{m_+^2}{\mu^2}\right) 
  -  \frac{m_\Psi^2}{2} \left(\frac{1}{\hat{\epsilon}} - {\rm Ln}\frac{m_-^2}{\mu^2}\right) \right] \\
 &=& - 2 \lambda^2 \frac{i}{16 \pi^2} \left[  \frac{m_+^2}{2}  {\rm Ln}\frac{m_+^2}{\mu^2} + 
 \frac{m_-^2}{2}  {\rm Ln}\frac{m_-^2}{\mu^2} - m_\Psi^2 {\rm Ln}\frac{m_\Psi^2}{\mu^2} \right. \nn \\
 && ~~~~~~~~~~~~~ \left.   + \frac{m_\Psi^2}{2}  {\rm Ln}\frac{m_+^2}{\mu^2} + 
 \frac{m_\Psi^2}{2}  {\rm Ln}\frac{m_-^2}{\mu^2} - m_\Psi^2 {\rm Ln}\frac{m_\Psi^2}{\mu^2}  \right] \\
&=& - 2 \lambda^2 \frac{i}{16 \pi^2} \left[  m_\Psi^2  {\rm Ln}\frac{m_+ \, m_-}{m_\Psi^2} + 
\lambda \langle F_x \rangle {\rm Ln}\frac{m_+}{m_-} + m_\Psi^2  {\rm Ln}\frac{m_+ \, m_-}{m_\Psi^2}
\right]
\end{eqnarray}
Hence, assuming $\Phi_1$ ($\Phi_2$) to be a ${\bf 5}$ ($\bf \bar{5}$) of $SU(5)$, and for $N_m$ pairs of $\{\Phi_1,\Phi_2\}$, we have, 
\bea
\Pi(p^2=0) &=&  5\, N_m\, \frac{2 \, \lambda^2}{16 \pi^2} \left[  2m_\Psi^2  {\rm Ln}\frac{m_+ \, m_-}{m_\Psi^2}  + 
\lambda \langle F_x \rangle {\rm Ln}\frac{m_+}{m_-} \right] \\
&=& 5 \, N_m \, \frac{\lambda^2}{16 \pi^2} m_\Psi^2 \left[  2  {\rm Ln}\frac{m_+^2 \, m_-^2}{m_\Psi^4}  + 
\frac{\lambda \langle F_x \rangle}{m_\Psi^2} {\rm Ln}\frac{m_+^2}{m_-^2} \right] \\
&=&  \left(4 \, \pi \, \sqrt{5 \, N_m} \right)^2 \,  \left(\frac{\lambda}{16 \pi^2}\right)^2 \frac{\lambda^2 \fvev^2 }{m_\Psi^2} G\left(\frac{\lambda \fvev}{m_\Psi^2}\right)
 \nn \\
\eea
where the function $G(x)$ is given by, 
\bea
G(x) = \frac{1}{x^2}[(2+x){\rm Log}(1+x) + (2-x){\rm Log}(1-x)] \, .
\eea

In terms of gaugino mass, this can be written as,  
\bea
\Pi(p^2=0) &=&   - \left(\frac{\lambda}{g_a^2}\right)^2 \, \left(4 \pi \, \sqrt{\frac{5}{N_m}} F(x)\right)^2 \, m_a^2 
\eea

The behaviour of the function $F(x) \equiv \sqrt{-G(x)/g(x)^2}$ is shown in Fig.~\ref{fig:Fx}.

\begin{figure}[h]
\centering
\begin{tabular}{c}
\includegraphics[scale=0.6]{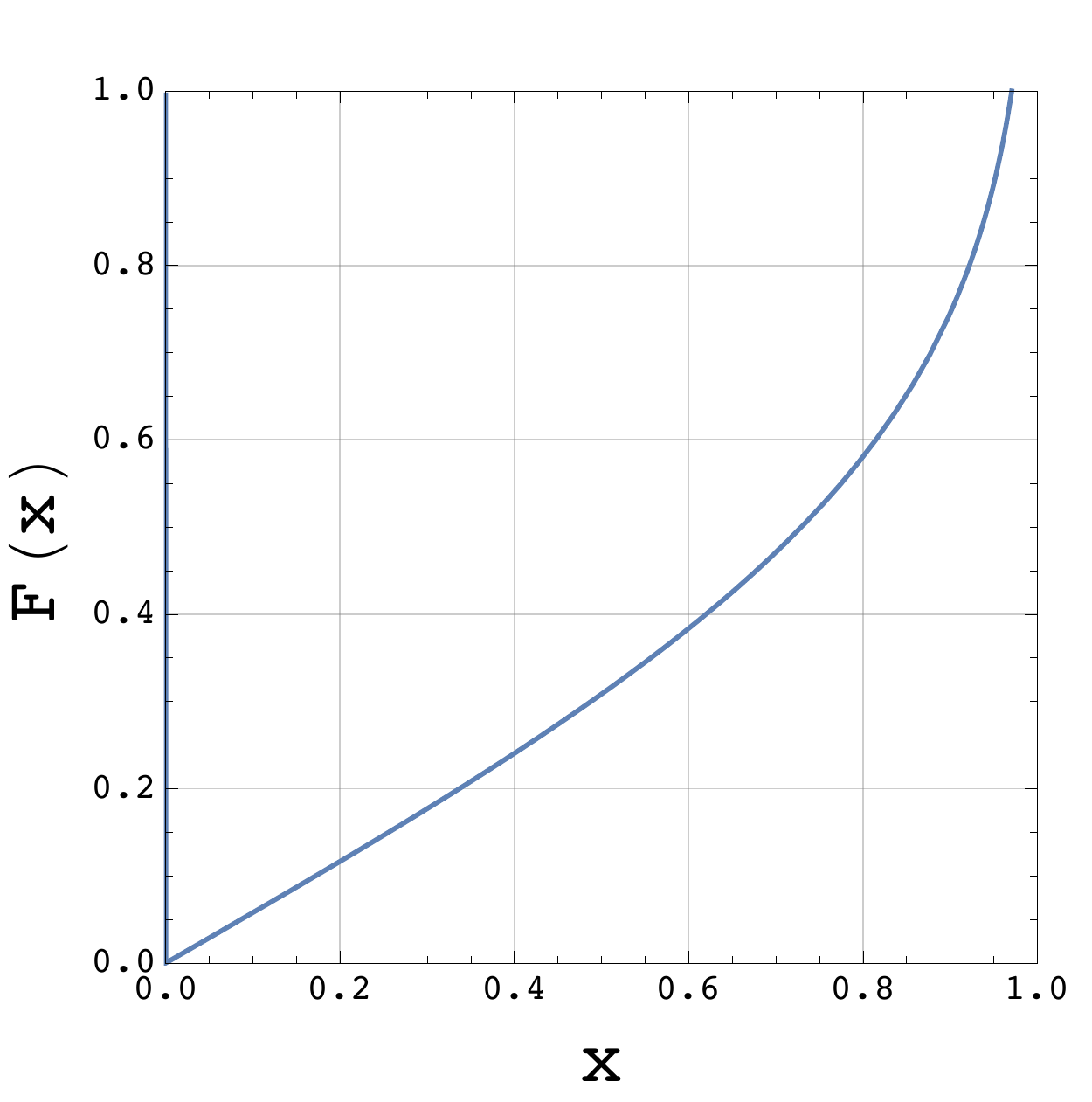}
\end{tabular}
\caption{{\sf The behaviour of $F(x)$ as a function of $x$.}}
\label{fig:Fx}
\end{figure}

\subsection{Coleman-Weinberg potential}

The Dirac mass for the fermions as a function of ${\cal S}$ is given by, 
\bea
m_F({\cal S})=\lambda {\cal S}
\eea
and the scalar mass matrix is 
\begin{eqnarray}
{\tilde m}^2 ({\cal S}) = \left(\begin{array}{cc}
\lambda^2 {\cal S}^* {\cal S} & - \lambda \fvev \\
- \lambda \fvev & \lambda^2 {\cal S}^* {\cal S}
\end{array}
 \right) \, ,
\end{eqnarray}
with the eigenvalues,
\bea
m^2_\pm ({\cal S}^* {\cal S}) = \lambda^2 {\cal S}^* {\cal S} \pm \lambda \fvev \, .
\eea
Using the standard formula for the Coleman-Weinberg potential\cite{Coleman:1973jx},
\bea
V_{\rm CW}= \frac{1}{64 \pi^2}\, {\rm STr}\left( {\cal M}^4 \left[ \log 
\frac{{\cal M}^2}{\Lambda^2_{\rm cut-off}} -\frac32 \right] \right),
\eea
we get,
\bea
V_{\rm CW}&=&\frac{{ 2}}{64 \pi ^2} \Big[
\left[m_+^2({\cal S},{\cal S}^*)\right]^2  {\rm Ln} \left[m_+^2({\cal S},{\cal S}^*)\right] 
+\left[m_-^2({\cal S},{\cal S}^*)\right]^2  {\rm Ln} \left[m_-^2({\cal S},{\cal S}^*)\right] \nn \\
&& ~~~ -2 \left[m_F({\cal S})^* \, m_F({\cal S})\right]^2 {\rm Ln} \left[ m_F({\cal S})^* \, m_F({\cal S}) \right] \nn \\
&& ~~~ -  \lambda^2 \fvev^2 \left( 
\log\Lambda^2_{\rm cut-off} + \frac32\right) \Big]
\eea

After replacing ${\cal S} \to \svev + {\cal S}$, we get the coefficient of ${\cal S}^* {\cal S}$ to be, 
\bea
\Pi(p^2=0) &=&  \frac{2 \, \lambda^2}{16 \pi^2} \left[  2m_\Psi^2  {\rm Ln}\frac{m_+ \, m_-}{m_\Psi^2}  + 
\lambda \langle F_x \rangle {\rm Ln}\frac{m_+}{m_-} \right]
\eea

\subsection{Tree level sgoldstino mass}
\label{tree-sgoldstino}
Here we give an example of a model where the sgoldstino gets tree level mass at the time of SUSY breaking \cite{Caracciolo:2012de}. 
The mode is just an extension of  the Affleck-Dine-Seiberg model (ADS) or 3-2 model of \cite{Affleck:1984xz,Peskin:1997qi,Poppitz:1998vd}. 
The field content of the ADS model is 
\begin{center}
\begin{tabular}{ccc}
\hline
 & $\SU{3}$ & $\SU{2}$ \\
 \hline
$Q$ & $3$ & $2$\\
$U^c$ & $\overline{3}$ & $1$\\
$D^c$ & $\overline{3}$ & $1$\\
$L$ & $1$ & $\overline{2}$\\
 \hline
\end{tabular},
\end{center}
and the superpotential is given by, 
\begin{eqnarray}
W_{3-2} &=& W_{{\rm cl}} + W_{{\rm np}} \, \, \\ 
\rm where,  && \, \\
W_{{\rm cl}} &=& h Q_A^a D_a^c L^A \, , \\
W_{{\rm np}} &=& \frac{\Lambda_3^7}{{\rm det}(Q Q^c)}\, ,
\end{eqnarray}
where, $Q^c$ is defined as $Q^c \equiv (U^c , D^c)$.
In this model $h << \tilde g_2,\tilde g_3$ which are the gauge couplings for the groups $\SU{2}$ and $\SU{3}$ respectively. 
Thus, $F$-term contribution to the scalar potential is subdominant compared to the $D$-term contribution. 
The minimum of the potential can be obtained perturbatively along the $D$-flat directions, 
\begin{eqnarray}
Q = \left( \begin{array}{cc}
a & 0 \\
0 & b \\ 
0 & 0 
\end{array}\right) M \ \ \ ,  \ \ \  Q^c = \left( \begin{array}{cc}
a & 0 \\
0 & b \\ 
0 & 0 
\end{array}\right) M \ \ \ , \ \ \ L = \left( \begin{array}{cc}
\sqrt{a^2 - b^2} \\
0
\end{array}\right) M
\end{eqnarray}
where,
\beqa
M \equiv \frac{\Lambda_3}{h^{1/7}} \gg \Lambda_3 \, ,
\eeqa
and $a\approx 1.164$, $b\approx 1.132$.

Note that $L_1$ (the component of $L$ getting a non-vanishing VEV) is the sgoldstino here. The $\SU{2}$ $D$-term equation of motion gives, 
\begin{equation}
\label{3-2-Dterm}
D^a_2  = {\tilde g}_2 \sum_f f^\dag T^a_2 f 
\end{equation}
where  $T^a_2 = \sigma^a/2$, $\sigma^a$ being the Pauli matrices.  
The Eq.~\ref{3-2-Dterm} will get contributions from all the fields carrying $\SU{2}$ charge i.e., $Q$ and $L$,
\begin{equation}
D^a = \tilde g_2 \left( L^\dag \frac{\sigma^a}{2} L + \sum_r Q^{r \dag}_i 
\frac{\sigma^a}{2} Q^r_i \right) 
\end{equation}
where the index $r$ is the $\SU{3}$ index. This gives, for the scalar potential, 
\begin{eqnarray}
V &=& \frac{1}{2} D^a D_a \\
&=& \frac{\tilde g_2^2}{8} \left( L^\dag \frac{\sigma^a}{2} L + \sum_r Q^{r 
\dag}_i \frac{\sigma^a}{2} Q^r_i \right) \left( L^\dag \frac{\sigma_a}{2} L + 
\sum_r Q^{r \dag}_i \frac{\sigma_a}{2} Q^r_i \right) \, .
\end{eqnarray}
Noting that only the third Pauli matrix contributes, we have, 
\begin{eqnarray}
V&=& \frac{\tilde g_2^2}{8} \left[\left(L_1^\dag L_1\right)^2 + 2 (L_1^\dag 
 L_1)(Q^{r\dag}_1 Q^r_1 -Q^{r\dag}_2 Q^r_2) + \cdots\right],
\end{eqnarray}
where the ellipses denote terms unimportant for sgoldstino mass. 
This generates a mass term for $L_1$ which is given by, 
\begin{eqnarray}
M_{L_1}^2 &=& \frac{\tilde g_2^2}{8} \left(4 (a^2 - b^2) M^2 + 2 (a^2 - b^2) M^2\right) \\
&=& \frac{3 \tilde g_2^2}{4} (a^2 - b^2) M^2 \, .
\end{eqnarray}
This is, in general, much larger than the gaugino mass. 

\section{Calculation of the gaugino mass }

The relevant part of the Lagrangian is given by 
\begin{equation}
\mathcal{L} \subset  \int d^4\theta \, \Phi_1^\dag e^{2g T^a V^a} \Phi_1 + \int d^4\theta \, \Phi_2^\dag e^{ 2g T^a V^a} \Phi_2  
+ \left( \int d^2 \theta \, y X \Phi_1 \Phi_2  + \rm h.c. \right)
\end{equation}
where,
\begin{equation}
V^a = \theta \bar{\sigma}^\mu \bar{\theta} A_\mu^a + i \theta^2 \bar{\theta} \lambda^{\dag a} - i \theta \bar{\theta}^2  \lambda^a + \frac{1}{2} \theta^2 \bar{\theta}^2 D^a
\end{equation}

A Majorana fermion $\Psi_\lambda$ is constructed out of the (Weyl) gaugino field $\lambda_a$, 
\bea
\Psi_\lambda = \left( 
\begin{array}{c}
\lambda_{\alpha} \\ 
\lambda^{\dagger \dot \alpha}\end{array}
\right)
\eea

\begin{figure}[H]
\begin{center}
\includegraphics[scale=0.85]{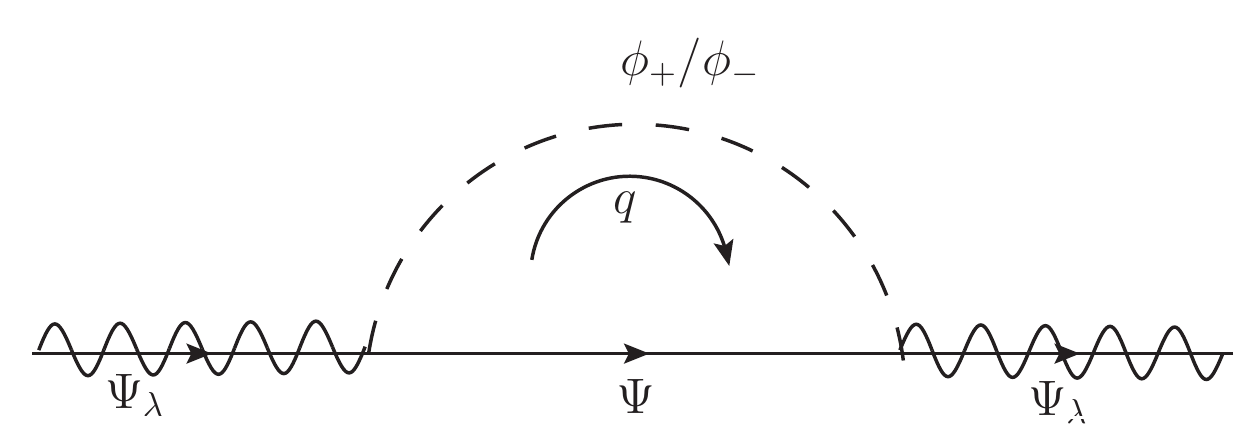}
\end{center}
\caption{{\sf One-loop contribution to the gaugino mass.}}
\label{gaugino_fig}
\end{figure}
 
The relevant vertex factors are given by, 
\begin{eqnarray}
\alpha \   \overline{\Psi}_\xi \Psi^A_\lambda \ \ &:& \ \ -i g T^A \\
\beta \   \overline{\Psi}_\xi \Psi^A_\lambda \ \ &:& \ \ +ig \gamma_5 T^A
\end{eqnarray}

The gaugino mass is generated via the one loop diagrams shown in Fig.~\ref{gaugino_fig}.

\subsubsection*{Loop with the scalar $\alpha$} 
\begin{eqnarray}
-\frac{i}{2}m_{\Psi_\lambda}^{(\alpha)AB} &=&  {\rm Tr}[T^A T^B]\int \frac{d^4q}{(2\pi)^4} \left(-g\right)  \frac{-i}{\not{\!q}+m_\Psi} \left(+g\right) \frac{i}{q^2 - m_\alpha^2} \\
&=&  -g^2  \, {\rm Tr}[T^A T^B] \int \frac{d^4q}{(2\pi)^4} \frac{\not{\!q} - m_\Psi}{q^2 - m_\Psi^2} \frac{1}{q^2 - m_\alpha^2} \\
&=& g^2 m_\Psi  \, {\rm Tr}[T^A T^B] \int \frac{d^4 q}{(2\pi)^4} \frac{1}{q^2 - m_\Psi^2} \frac{1}{q^2 - m_\alpha^2} \\
&=& g^2 m_\Psi  \, {\rm Tr}[T^A T^B] \, B_0(0,m_\Psi,m_\alpha) 
\end{eqnarray}
\subsubsection*{Loop with the scalar $\beta$} 
\begin{eqnarray}
-\frac{i}{2}m_{\Psi_\lambda}^{(\beta) AB}  &=& {\rm Tr}[T^A T^B] \int \frac{d^4q}{(2\pi)^4} \left(-g \gamma_5\right)  \frac{-i}{\not{\!q}+m_\Psi} 
\left(-g \gamma_5\right) \frac{i}{q^2 - m_\beta^2} \\
&=& g^2 {\rm Tr}[T^A T^B]  \int \frac{d^4q}{(2\pi)^4} \gamma_5 \frac{\not{\!q} - m_\Psi}{q^2 - m_\Psi^2} \gamma_5 \frac{1}{q^2 - m_\beta^2} \\
&=& g^2 {\rm Tr}[T^A T^B]  \int \frac{d^4q}{(2\pi)^4} \frac{(-\not{\! q}-m_\Psi)}{q^2 - m_\Psi^2} \frac{1}{q^2 - m_\beta^2}\\
&=& -g^2 m_\Psi  \, {\rm Tr}[T^A T^B] \, B_0(0,m_\Psi,m_\beta) 
\end{eqnarray}

where, the $B_0$ function is given by, 
\bea
B_0(0,m_1,m_2) = \frac{A_0(m_1) - A_0(m_2)}{m_1^2 - m_2^2}
\eea

\begin{eqnarray}
-\frac{i}{2}m_{\Psi_\lambda}^{AB}  &=&  g^2 m_\Psi  \, {\rm Tr}[T^A T^B] \, (B_0(0,m_\Psi,m_\alpha) - B_0(0,m_\Psi,m_\beta)) \\
&=&  -i\frac{g^2 m_\Psi}{16 \pi^2}  \, {\rm Tr}[T^A T^B] \, \frac{\left(1 + x\right) \ln \left(1+x\right) + \left(1 - x\right) \ln \left(1-x\right)}{x} \\
&=&  -i\frac{g^2}{16 \pi^2}  \, {\rm Tr}[T^A T^B] \, \frac{\fvev}{m_\Psi} \, \frac{\left(1 + x\right) \ln \left(1+x\right) + \left(1 - x\right) \ln \left(1-x\right)}{x^2} \\
m_{\Psi_\lambda}^{AB}&=&  \frac{g^2}{16 \pi^2}  \,  2{\rm Tr}[T^A T^B] \, \frac{\fvev}{m_\Psi} \, g(x) \\
m_{\Psi_\lambda}^{AB}&=&  \frac{g^2}{16 \pi^2}  \,  \frac{\fvev}{m_\Psi} \, g(x) \, \delta^{AB} \, .
\end{eqnarray}

%
\providecommand{\href}[2]{#2}\begingroup\raggedright\endgroup

%
\end{document}